\documentclass[prd,twocolumn,superscriptaddress,showpacs,floatfix,longbibliography,nofootinbib]{revtex4-1}
\usepackage[utf8]{inputenc}
\usepackage{graphicx,amsfonts,amssymb,amsmath,xspace,physics}
\usepackage{comment}
\usepackage[pdftex,breaklinks=true,colorlinks=true]{hyperref}
\usepackage{epstopdf}
\usepackage{tabularx}
\usepackage{array}
\usepackage{multirow}
\usepackage{diagbox}
\usepackage{enumitem}
\usepackage{xcolor,colortbl}
\usepackage{stmaryrd}
\definecolor{g}{rgb}{.1,0.4,.1} 
\definecolor{b}{rgb}{0,0.2,1}
\definecolor{rouge}{rgb}{0.82,0.,0.}
\definecolor{vert}{rgb}{0.,0.82,0.}
\definecolor{orange}{rgb}{1,0.5,0.}
\definecolor{bleu}{rgb}{0.,0.,0.82}
\definecolor{m}{rgb}{0.82,0.,0.82}
\definecolor{vert2}{rgb}{0.,0.5,0.}
\definecolor{rougeclair}{rgb}{1.0,0.7,0.7}
\definecolor{gris}{rgb}{0.8,0.8,0.8}
\newcolumntype{g}{>{\columncolor{gris}}m}

\newcommand{\beq}{\begin{equation}}
\newcommand{\be}{\begin{equation}}
\newcommand{\beqn}{\begin{eqnarray}}
\newcommand{\eeq}{\end{equation}}
\newcommand{\ee}{\end{equation}}
\newcommand{\eeqn}{\end{eqnarray}}

\newcommand{\bem}{\begin{pmatrix}}
\newcommand{\eem}{\end{pmatrix}}

\newcommand{\f}{\frac}

\newcommand{\e}{\textrm{e}}
\newcommand{\im}{\textrm{i}}

\newlength{\ldag}
\begin{document}
\title{Robustness of Aharonov-Bohm cages in quantum walks}

\author{Hugo Perrin}
\email{hugo.perrin@kit.edu}
\affiliation{Sorbonne Universit\'e, CNRS, Laboratoire de Physique Th\'eorique de la Mati\`ere Condens\'ee, LPTMC, 75005 Paris, France}
\affiliation{Karlsruhe Institute of Technology, 76021, Karlsruhe, Germany}

\author{Jean-No\"el Fuchs}
\email{fuchs@lptmc.jussieu.fr}
\affiliation{Sorbonne Universit\'e, CNRS, Laboratoire de Physique Th\'eorique de la Mati\`ere Condens\'ee, LPTMC, 75005 Paris, France}

\author{R\'emy Mosseri}
\email{remy.mosseri@upmc.fr}
\affiliation{Sorbonne Universit\'e, CNRS, Laboratoire de Physique Th\'eorique de la Mati\`ere Condens\'ee, LPTMC, 75005 Paris, France}

\date{\today}

\begin{abstract}
It was recently shown that Aharonov-Bohm (AB) cages exist for quantum walks (QW) on certain tilings -- such as the diamond chain or the dice (or $\mathcal{T}_3$) lattice -- for a proper choice of coins. In this article, we probe the robustness of these AB cages to various perturbations. When the cages are destroyed, we analyze the leakage mechanism and characterize the resulting dynamics. Quenched disorder typically breaks the cages and leads to an exponential decay of the wavefunction similar to Anderson localization. Dynamical disorder or repeated measurements destroy phase coherence and turn the QW into a classical random walk with diffusive behavior. Combining static and dynamical disorder in a specific way leads to subdiffusion with an anomalous exponent controlled by the quenched disorder distribution. Introducing interaction to a second walker can also break the cages and restore a ballistic motion for a ``molecular'' bound-state. 
\end{abstract}


\maketitle
\date{\today}

\section{Introduction}
\label{sec:intro}
Aharonov-Bohm (AB) cages were discovered in the Hofstadter butterfly of the $\mathcal{T}_3$ lattice~\cite{Vidal98}. It is a strict confinement phenomenon that occurs for a single electron described by a tight-binding model on a certain 1D or 2D periodic lattice containing loops and subjected to a perpendicular magnetic field. The extreme localization is due to destructive AB interferences occurring at a critical magnetic flux, usually corresponding to half a flux quantum per plaquette. The wavefunctions corresponding to cages are of finite support and not exponentially localized, as in Anderson localization by disorder, for example. The energy spectrum at the critical flux consists of completely flat bands. When varying the magnetic field around the critical value, one observes a characteristic pinching of the energy bands as they become flat.  
AB cages were studied in several experiments with superconducting wire networks \cite{Abilio99}, Josephson junctions arrays \cite{Pop2008}, cold-atomic gases \cite{moller12}, photonic lattices \cite{Mukherjee18} and ion microtraps \cite{Porras2011}.

Recently, it was shown that AB cages can also occur in the context of discrete-time quantum walks (QW) and are not restricted to tight-binding Hamiltonians~\cite{Perrin2020}. In short, QW are basically unitary transformations on a graph for a particle with internal degree of freedom, the state of the latter deciding which direction (edge of the graph) is taken at the next step. At each time step a unitary transformation (a ``quantum coin" or simply a coin) is applied to the internal state of the walker. For AB cages to happen in such a system, one needs to find a proper tiling (e.g. diamond chain or $\mathcal{T}_3$ or $\mathcal{T}_4$ tilings) and a specific set of coins that are able to realize such a destructive interference tuned by the magnetic field. The cages found are similar to the Hamiltonians ones but have marked differences such as varying sizes and the possibility to tune the critical flux by playing with the coin. In addition, due to the unitary nature of the transformations, the corresponding spectrum versus magnetic field has a Floquet nature and is periodic not only in flux but also in quasi-energy (later defined from the overall unitary transformation spectrum). 

In the present article, we concentrate on the simplest example of AB cages for the QW on the diamond chain (DC) and ask the following question: how robust are these QW AB cages when perturbed by disorder, measurements or interactions? Similar questions were studied in the Hamiltonian context in~\cite{Vidal2000,Vidal2001}. It was generically found that a second on-site interacting particle or a disordered on-site potential destroy the AB cages.

In the present QW context, whenever AB cages are destroyed, we characterize the resulting leakage dynamics, which leads to several other questions: Do we recover the ballistic motion characteristic of usual QW or diffusive behavior as in the classical random walks? Does disorder lead to Anderson-like localization? Is it possible to obtain more exotic dynamics upon destroying the cages? We answer these questions in the present work.

The article is organized as follows. We first review, for the sake of completeness, the QW on the diamond chain in Sec.~\ref{sec:QWDC} before introducing various perturbations. In Sec.~\ref{sec:statdis}, we study the effect of static disorder. Then, in Sec.~\ref{sec:dyndis}, we consider dynamical disorder and study how it destroys phase coherence. In Sec.~\ref{sec:measure}, we show that repeated projective measurements have a similar decoherence effect. In the following Sec.~\ref{sec:subdiffusion}, we combine static and dynamic disorder in a specific way such as to produce anomalous diffusion with a non-trivial exponent controlled by the coin. In Sec.~\ref{sec:2ndparticle}, we study the effect of interaction on the AB cages by introducing a second quantum walker. Finally in Sec.~\ref{sec:conclusion}, we conclude and give perspectives. Calculation details are provided in several Appendices.

\section{Quantum walk on the diamond chain } 
\label{sec:QWDC}
\begin{figure}[t]
    \includegraphics[width=0.5\textwidth]{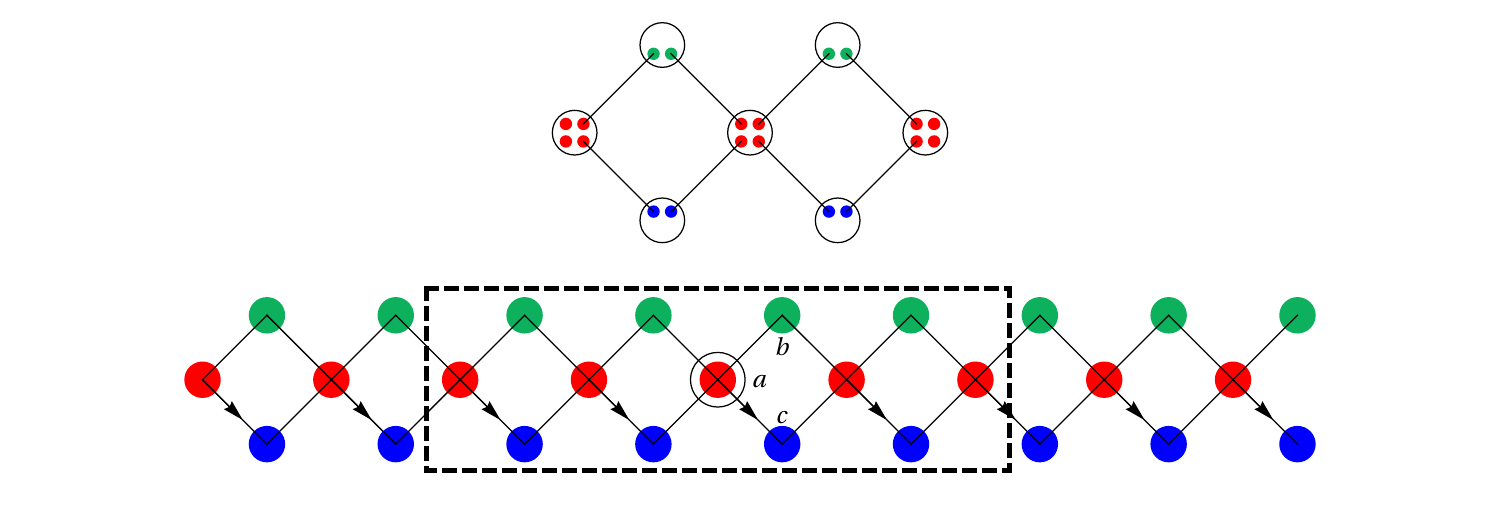}
\caption{Bottom: diamond chain with colored sites $a$ (red), $b$ (green) and $c$ (blue). Arrows indicate the phase factor $e^{i 2\pi f}$. The dashed rectangle is the maximal extension of a cage at $f_c$, for an initial state localized on the circled $a$ site. Top: the Hilbert space is schematized with four (resp. two) basis states for $a$ (resp. $b,c$) sites, shown here as circles. Coins operate on states inside a circle, and shifts along  edges. Figure taken from~\cite{Perrin2020}.}
    \label{fig:diamond chain}
\end{figure}

In this section, we sum up the basic lines and the important results  of~\cite{Perrin2020}. We consider a QW in a perpendicular magnetic field and study cage effects on the DC with four-fold coordinated ``hub'' sites $a$ and two-fold ``rim'' sites $(b,c)$ (see Figure~\ref{fig:diamond chain}). To define a QW on such lattice, we equipped sites with internal states encoding the direction of the walker. Their internal space's dimension is equal to their connectivity (see top of Figure~\ref{fig:diamond chain}). In each unit cell, the internal space size is $8$. The shift operator $S$ connects every  internal states sharing the same edge. It is unitary and hermitian and reads:
\begin{equation}
S=\sum_{\left\langle( i,j),(i',j')\right\rangle}\ket{i,j}\bra{i',j'} +\text{h.c.},
\end{equation}
 where $i$ and $i'$ are neighbouring sites, and $j$ and $j'$ are the two internal states connecting these sites.

The operator $C_s$ operates a linear combination of the internal states belonging to the same site $s=\{a,b,c\}$. The coin operator $C$ is the direct sum of $C_a, C_b$ and $C_c$.
We need to define two types of coin associated with either rim or hub sites. For the two-fold (rim) sites, we use generic unitary  coins:
\begin{equation}
\label{eq:coin}
U_2(\theta,\varphi,\omega,\beta)=
\begin{pmatrix}
\cos \theta \, {\rm e}^{\im  \beta}& -\sin \theta \, {\rm e}^{\im (\varphi+\omega)} \\
\sin \theta \, {\rm e}^{-\im  \omega} &\cos\theta \, {\rm e}^{i(\varphi-\beta)}
\end{pmatrix}.
\end{equation}
 For hub sites, we use either  $H_4=H_2\otimes H_2$, where $H_2$ denotes the $2\times 2$ Hadamard matrix or $G_4 =\frac{1}{2} \,\mathbf{1}_4- \mathbb{I}_4$, where $\mathbf{1}_4$ is the $4\times 4$ matrix full of $1$ and $\mathbb{I}_4$ is the identity matrix. $G_4$ is the Grover coin of dimension $4$. We use the same basis conventions for the matrix representation of coins as in~\cite{Perrin2020}.
 
 \par  The quantum walk operator is then the product of both operations $W=SC$. $W$ being unitary, its eigenvalues are pure phases, called quasi-energies and defined modulo $2\pi$.
 
\par The magnetic field \mbox{$B=|\boldsymbol{\nabla}\times \boldsymbol{A}|$} enters via a Peierls substitution~\cite{Peierls33}, i.e. the hopping terms in the shift $S$ get multiplied by a phase factor 
 $
{\rm e}^{\im \frac{2\pi}{\phi_0} \int_{i}^{i'} \textbf{dl}\cdot \boldsymbol{A}}
$, where $\boldsymbol{A}$ is the vector potential and $\phi_0=h/e$ the flux quantum~\cite{Yalcinkaya15, alberti19, Cedzich19}. 

For the DC, it is possible to find a gauge that preserves the structure periodicity. We choose it as a phase ${\rm e}^{\im 2\pi f}$ on one of the four edges (see Figure~\ref{fig:diamond chain}), with the reduced flux $f$ defined as the magnetic flux per plaquette in units of $\phi_0$. 
 
\par
\begin{figure}[h]

\begin{tabular}{c}
   \includegraphics[width=0.45\textwidth]{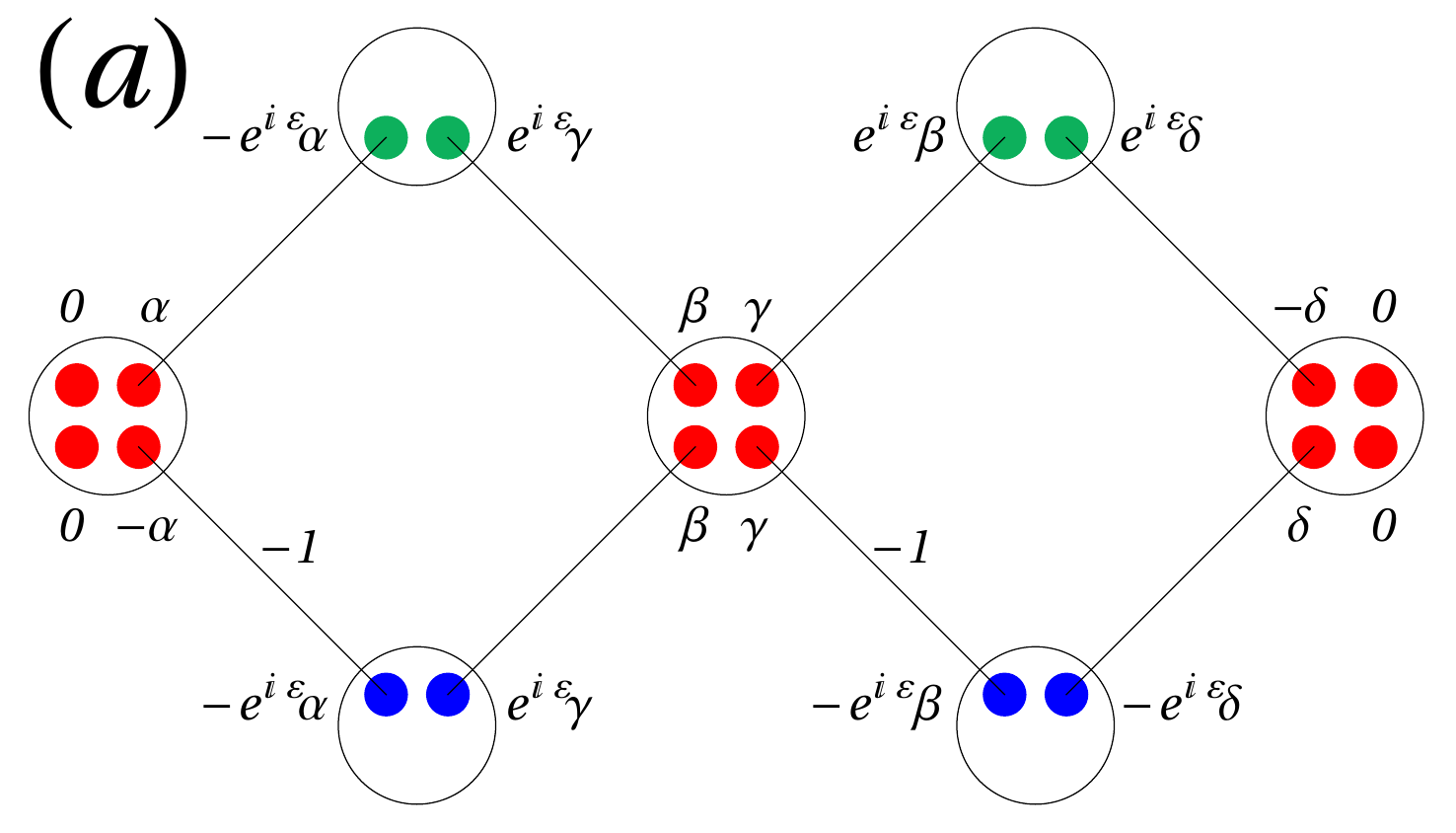}\\
   \includegraphics[width=0.45\textwidth]{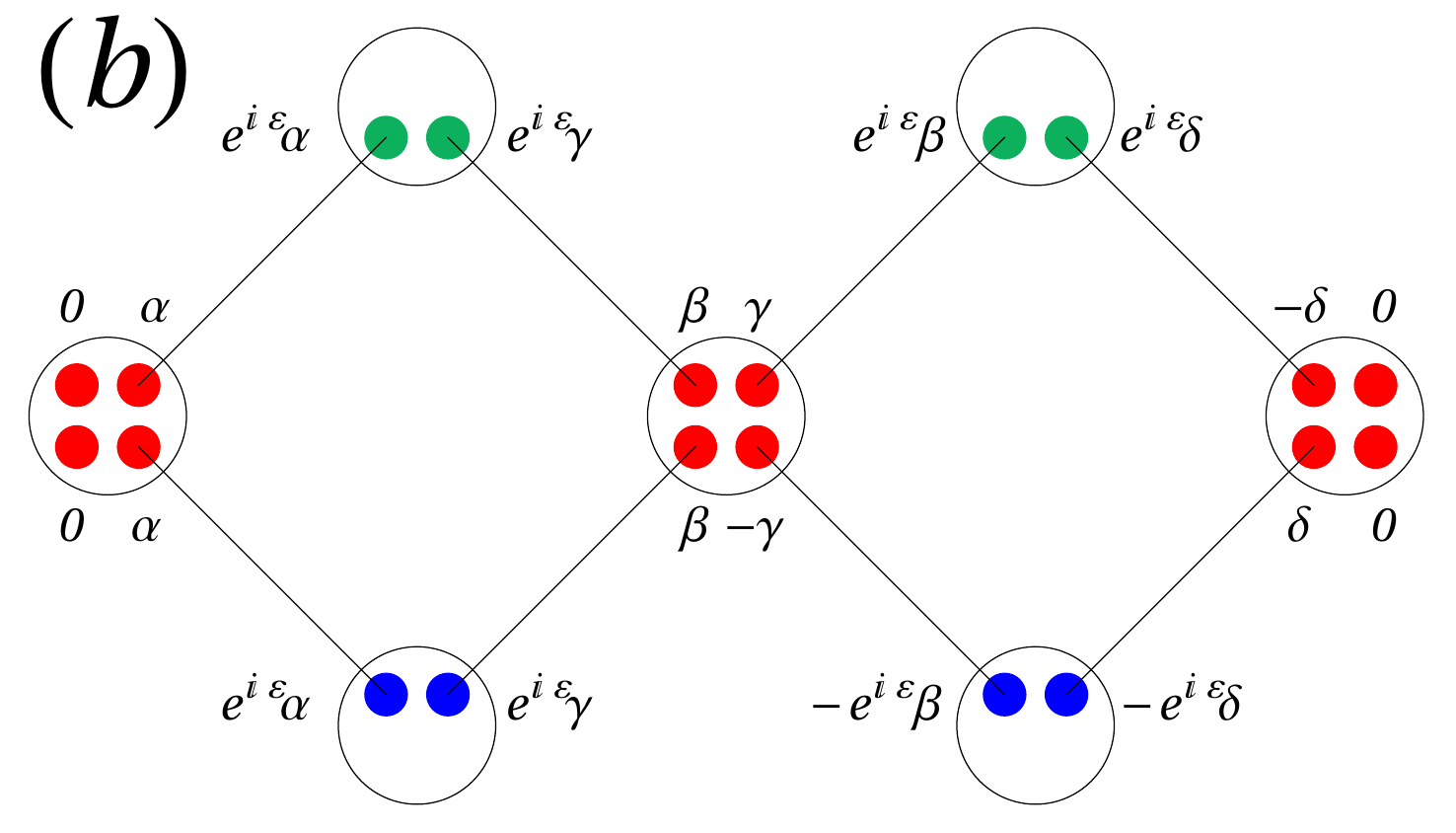}
\end{tabular}
  \caption{Maximally confined eigenvectors of QW on DC for (a) a Grover coin at $f_c=1/2$ (the gauge carries the Peierls phase on the edge connecting the $a$ sites to the right $c$ sites) and (b) a Hadamard coin at $f_c=0$. The coefficients $\alpha$, $\beta$, $\gamma$ and $\delta$ depend on the quasi-energy $\varepsilon$ as given in Table \ref{tab:coef}.}
    \label{fig:eigenvectors}

\end{figure}

 A detailed study of the spectrum is provided in \cite{Perrin2020}. We observe at the critical flux $f=1/2$ (resp. $f=0$) for Grover coins (resp. Hadamard coins) a pinching of the quasi-energy associated with an AB-like caging effect. Each quasi-energy being highly degenerated, they correspond to an extensive number of eigenvectors. We choose to represent these eigenvectors in the basis which minimize their extension. We call them maximally confined eigenstates and they are displayed in Figure~\ref{fig:eigenvectors} for $\theta=\pi/4$, $\omega=\beta=\varphi=0$. Since they extend over 3 cells, an initial localised state has only weight on a limited number of maximally confined eigenstate, the quantum walker is, then, trapped on a region of finite size given by the extension of these eigenstates. We denote by $\ket{i,\varepsilon}$ those states where $i$ is the index cell of the hub site around which is centered the eigenstate and $\varepsilon$ is the corresponding energy of the eigenstate. 

\begin{table}
 
\begin{tabular}{|c||c|c|c|c|}

 \hline $\varepsilon$  & $\alpha$&$\beta$&$\gamma$&$\delta$\\
 \hline
 \hline \multicolumn{5}{|c|}{Grover}\\
\hline
  \hline  0, $\pi$ &$1$&$1+\sqrt{2}$&$1+\sqrt{2}$&1\\
    \hline  $\pm\pi/2$ &$1$&$1-\sqrt{2}$&$1-\sqrt{2}$&1\\
      \hline  $-3\pi/8$, $5\pi/8$ &$-1$&$i$&$-i$&1\\
      \hline  $3\pi/8$, $-5\pi/8$ &$-1$&$-i$&$i$&1\\
      \hline
      \hline \multicolumn{5}{|c|}{Hadamard}\\
      \hline
      \hline $5\pi/12$, $-7\pi/12$ &$-i \sqrt{5-2\sqrt{6}}$&$1 -(-1)^{1/6} \sqrt{2}$&$1 +(-1)^{5/6} \sqrt{2}$&1\\
      \hline $-\pi/12$, $11\pi/12$ &$-i (\sqrt{2} + \sqrt{3})$&$1 +(-1)^{1/6} \sqrt{2}$&$1 -(-1)^{5/6} \sqrt{2}$&1\\
      \hline $-5\pi/12$, $7\pi/12$ &$i \sqrt{5-2\sqrt{6}}$&$1 +(-1)^{5/6} \sqrt{2}$&$1 -(-1)^{1/6} \sqrt{2}$&1\\
      \hline $\pi/12$, $-11\pi/12$ &$i (\sqrt{2} + \sqrt{3})$&$1 -(-1)^{5/6} \sqrt{2}$&$1 +(-1)^{1/6} \sqrt{2}$&1\\
      \hline
\end{tabular}
  \caption{Coefficients of the (unnormalized) eigenvectors drawn in Figure \ref{fig:eigenvectors} for the different quasi-energy  of the Grover and Hadamard QW at $f_c=1/2,0$ respectively and rim coin $U_2(\pi/4,0,0,0)$.} 
  \label{tab:coef}
\end{table}

\par In Appendix G of~\cite{Perrin2020}, it was noted that starting from a chain where only one type of coin is used (either Hadamard $H_4$ or Grover $G_4$) and inserting  periodically on hub sites the other coin leads to an extension of the cage.  The greater the distance between two substitution coins is, the bigger is the cage. Sec.~\ref{sec:quenchedhubdis} is a generalisation of this result in which Hadamard and Grover coins are chosen randomly on each hub sites.

\section{Quenched disorder}
\label{sec:statdis}

In the tight-binding hamiltonian system on a DC, it has been shown ~\cite{Vidal2001} that a disordered on-sites potential leads to the destruction of the AB cages. We propose a detailed analysis of different disorder for QW AB cages and focus on disorder encoded in the coin operator. It should be pointed out that, unlike hamiltonian systems where the disorder introduced is hermitian, in QW systems, because we are working at the level of evolution operator, any disorder considered should keep the coin unitary in order to ensure the conservation of probability.  This restricts the choice of disorder. However, a complete study of the different disorder on the coin operator remains beyond the scope of this article. For disorder on hub sites we focus on a random choice between the two former coins used (Grover $G_4$ and Hadamard $H_4$ coins). For the disorder on rim sites, the analysis of the $2\times 2$ unitary coins is limited to the $SO(2)$ rotations group.

\subsection{Disorder on hub sites}
\label{sec:quenchedhubdis}

\par On sites $a$, we introduce disorder by choosing randomly between $G_4$ and $H_4$ for every sites. We, therefore, create a random chain using a Bernoulli distribution of probability $p_s$ to select $G_4$ and $\overline{p_s}=1-p_s$ to select $H_4$. We can compute explicitly the probability $p_f(n)$ to get a cage of size $n$ and then deduce the average extension of cages $\langle n(p_s)\rangle_f$ at the critical flux $f=0,\frac{1}{2}$. Whenever it remains finite, we claim that cages are not broken but have simply changed their size.

First, we recall the three rules, described in Appendix G of~\cite{Perrin2020}, used to determine
the extension of a cage starting from an initial state localized on a site $a$, in the DC unit cell labelled $n_0$, is: 
\begin{itemize}
\item  First rule : The coin on the initial site $n_0$ is irrelevant for the cage to occur. What counts are coins applied on neighbouring sites. 
\item  Second rule : On the right-hand side, the QW spreads until it meets a substitution coin which will stop it on the right next site. 
\item Third rule : On the left-hand side, the coin on the ﬁrst neighbour of the initial site does not matter. Then, the QW spreads and stops exactly when it meets a substitution coin.

\end{itemize}

\par To compute $p_f(n)$, we first look at the probability $p_f^L(n)$ (resp. $p^R_f(n)$) of the walker extending to $n$ hub sites to the left (resp. right) of the initial site  at the magnetic flux $f$. Using the second and third rule at $f=0$ (resp. $f=1/2$), we deduce that this is the probability of having an $H_4$ (resp. $G_4$) and $(n-2)G_4$ (resp. $H_4$) and we notice that $p_f^L(n)$=$p_f^R(n)$ so that
\begin{equation}
    p^{R/L}_0(n)=\overline{p_s}p_s^{n-2},\  p^{R/L}_{1/2}(n)=p_s\overline{p_s}^{n-2} .
\end{equation}
 The total extension of the cage is given by the probability: 
\begin{eqnarray}
    p_f(n)&=&\sum_{k=2}^{n-3}p^L_f(k)p^R_f(n-1-k) \\ 
    p_0(n)&=&\overline{p_s}^2p_s^{n-5}(n-4),\ p_{1/2}(n)=p_s^2\overline{p_s}^{n-5}(n-4)\nonumber\\ 
    &&\mathrm{for}\ n \ge 5. \nonumber
\end{eqnarray}
\par To go from $f=0$ to $f=1/2$, one just has to switch $p_s$ into $\overline{p_s}$.
\par Then, one can compute the average extension:
\begin{eqnarray}
\langle n(p_s)\rangle_f&=&\sum_{k=5}^\infty k p_f(k)\nonumber\\
\langle n\rangle_0&=&3+\frac{2}{\overline{p_s}}\ \mathrm{and}\ \langle n\rangle_{1/2}=3+\frac{2}{p_s}
\label{eq:avgext}
\end{eqnarray}

\par
\begin{figure}[h]
   \includegraphics[width=0.5\textwidth]{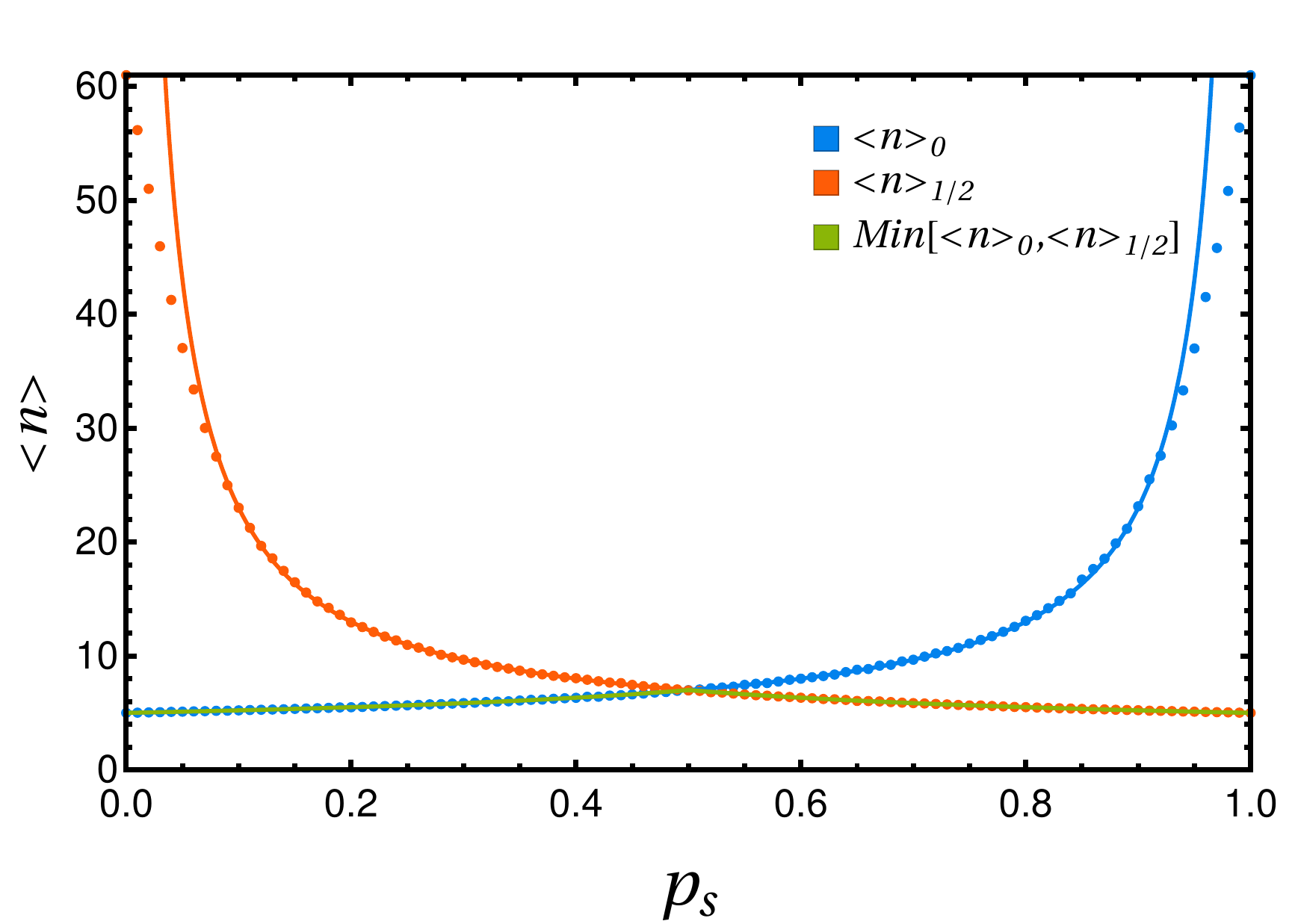}
  \caption{Mean value of the cage size with respect to the parameter $p_s$ for $f=0$ (blue curve) or $f=1/2$ (red curve). Solid lines are the predicted mean value (see Eq.~\eqref{eq:avgext}) while dotted lines represent the numerical simulation. Each point is an average over 10 000 QWs after 30 time steps. When the cage size is large the numerical simulation deviates from the theoretical results because of the finite time steps.}
    \label{fig:spahub}
\end{figure}

 The disordered system is characterized by two different coins, each associated with a different critical flux for caging. The average extension of the cages remains finite for both fluxes at all values of $p_s\in]0,1[$, i.e. cages resist this type of disorder. 
We plot on Figure~\ref{fig:spahub}, the Eq.~\eqref{eq:avgext} (solid line) and compare it to numerical simulations (dotted line). Both curves show a good agreement but deviates at $p_s,\bar{p_s}\ll 1$ (when cages size is large) due to the finite time steps of the simulation.

\par A domain of size $n$ of Grover (resp. Hadamard) is a location on the chain where we have a successive sequence of $n$ Grover (resp. Hadamard) coin bounded on each side by a Hadamard (resp. Grover) coin. When the DC size $L\to\infty$ and $p_s\neq0,1$, we expect to have Hadamard and Grover domains of all sizes. Suppose that we are at the flux $f=1/2$. The quantum walker can then freely move inside a Hadamard domain but its propagation is stopped by the Grover coins which bound the domain, these two coins acting as walls. Either the walker is stopped directly on the Grover coin for the left part of the quantum walk (third rule) or on the next hub site for the right part (second rule). When the walker starts in a Grover domain, at the flux $f=1/2$, it is trapped in the original QW AB cage. So we have two types of wave functions. Those whose extension is limited on 3 sites of the Grover domains, corresponding to caged eigenstates (see Figure~\ref{fig:eigenvectors}-a), with their 8 quasi-energies and those extended on a whole Hadamard domain with a marginal overlap on neighbouring Grover domains. Their corresponding quasi-energies are almost those of a quantum walker on a  diamond chain using Hadamard coin on the hub sites and with open boundary conditions. Their quasi-energies then depend on the size of the chain. For each Hadamard domain of size $n$ existing in the chain, there are $\simeq 8n$ associated quasi-energies. Several Hadamard domains can exist with the same size, this does not add any new quasi-energy in the spectrum but increases the degeneracy of the associated quasi-energies.

\par
\begin{figure}[h]
   \includegraphics[width=0.5\textwidth]{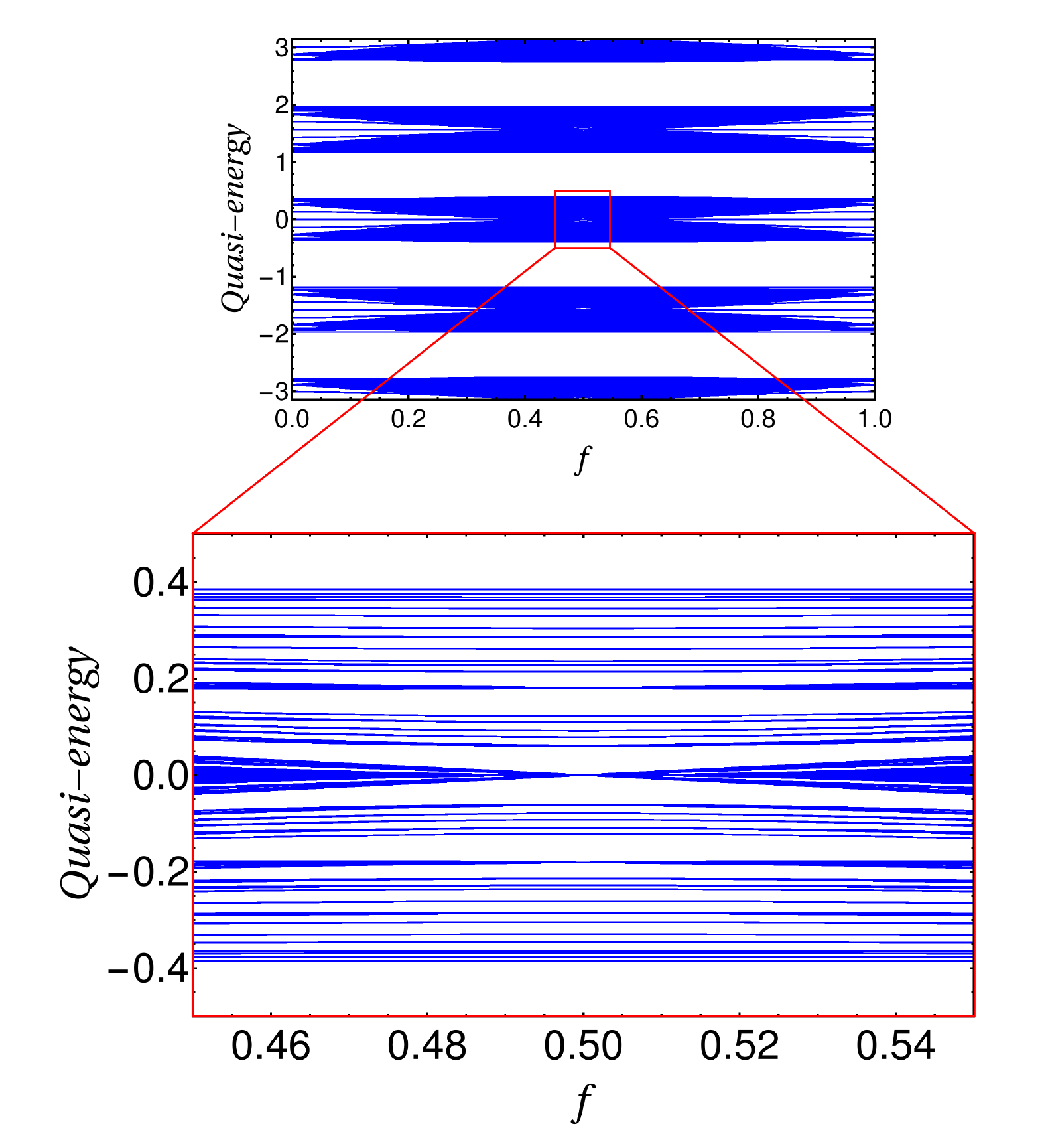}
  \caption{Spectrum of the QW for a quenched disorder on the hub sites with parameter $p_s=0.2$. Rim coins have $\theta=\pi/4$ and $\omega=\varphi=\beta=0$. Top: The whole quasi-energy $\varepsilon$ spectrum between $-\pi$ and $\pi$ as a function of the flux $f$ between $0$ and $1$. The spectrum seems to be dispersive at the critical flux $f=1/2$. Bottom: A zoom near $f=1/2$ and $\varepsilon=0$ (red square) reveals that the apparent continuous band is in fact composed of several flat bands at the critical flux.}
    \label{fig:spectrezoom}
\end{figure}

 Numerically, when we zoom in on the spectrum at flux $f=1/2$, we realize that the band which seemed to be dispersive is in fact composed of several bands which pinch at the critical flux (see Figure ~\ref{fig:spectrezoom}). One can of course do the same reasoning at $f=0$ by reversing the role of the Hadamard and Grover coin.

\subsection{Disorder on rim sites}
\label{sec:quenchedrimdis}

\par Let us now focus on the case where the quenched disorder is on the rim sites and in particular on the variable $\theta$ of the $2\times 2$ unitary matrix of the coin, see Eq.~(\ref{eq:coin}). The other parameters are taken to be $\varphi=\omega=\beta=0$. We choose a coin on the hub sites (either Hadamard or Grover) and set the flux at its critical value (either $f=0$ or $1/2$). 

\par If we take the parameter $\theta$ to be the same for the $b$ and $c$ sites of the same cell, the cages resist disorder, because the symmetry between the paths going through the sites $b$ and the site $c$ is maintained and therefore the destructive AB interferences are preserved.

\par We therefore consider a disorder such that the $\theta$ variables of the site $b$ and $c$ of a same cell are different, which unbalances the AB interferometer in each loop. We choose randomly the different $\theta$ variables in a box distribution $\left[\theta_0-\Delta\theta/2,\theta_0+\Delta\theta/2\right]$ where $\theta_0$ is the mean value and $\Delta\theta$ the strength of the disorder which can be chosen between $0$ and $2\pi$. 
\begin{figure}[h]
\centering
\begin{tabular}{cc}
 \includegraphics[width=0.25\textwidth]{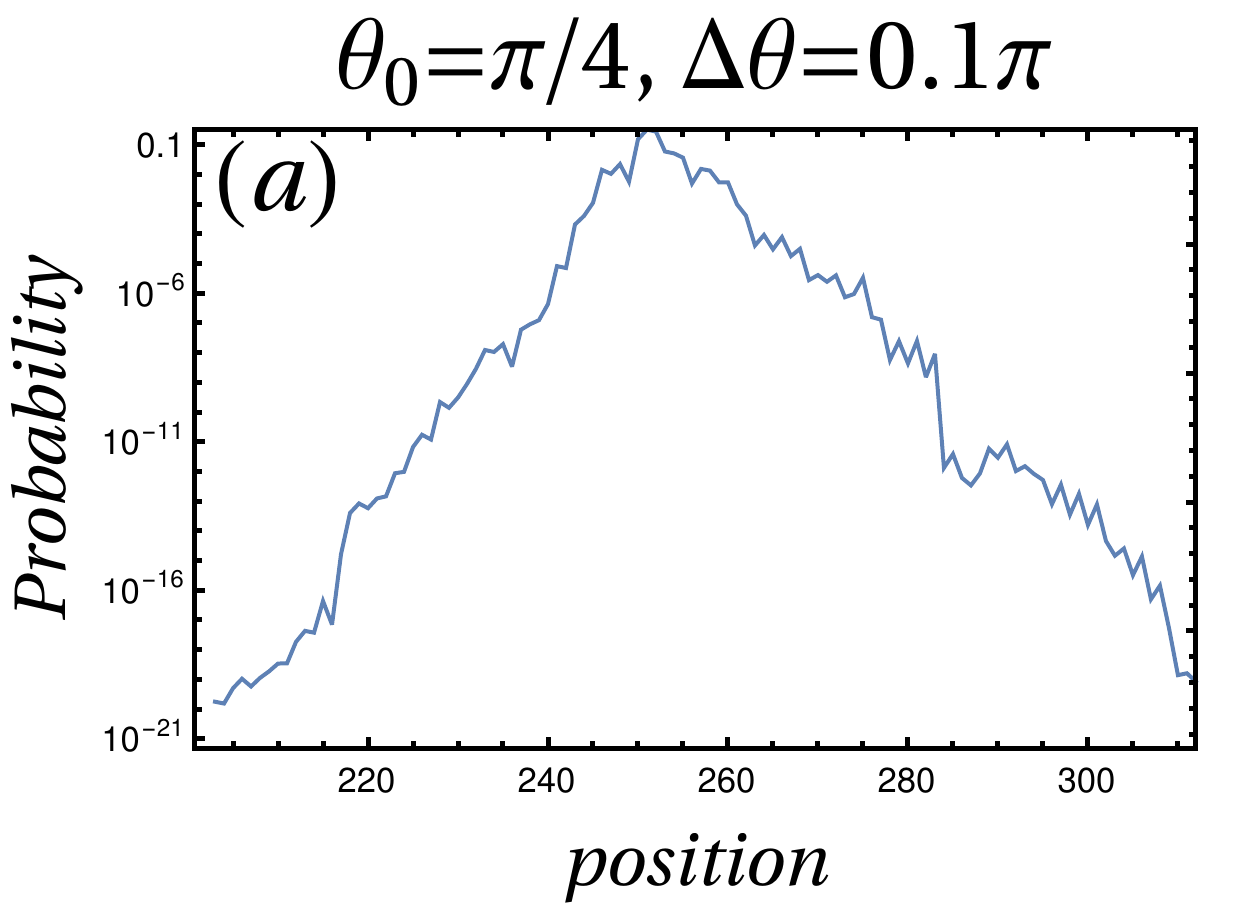}   & \includegraphics[width=0.25\textwidth]{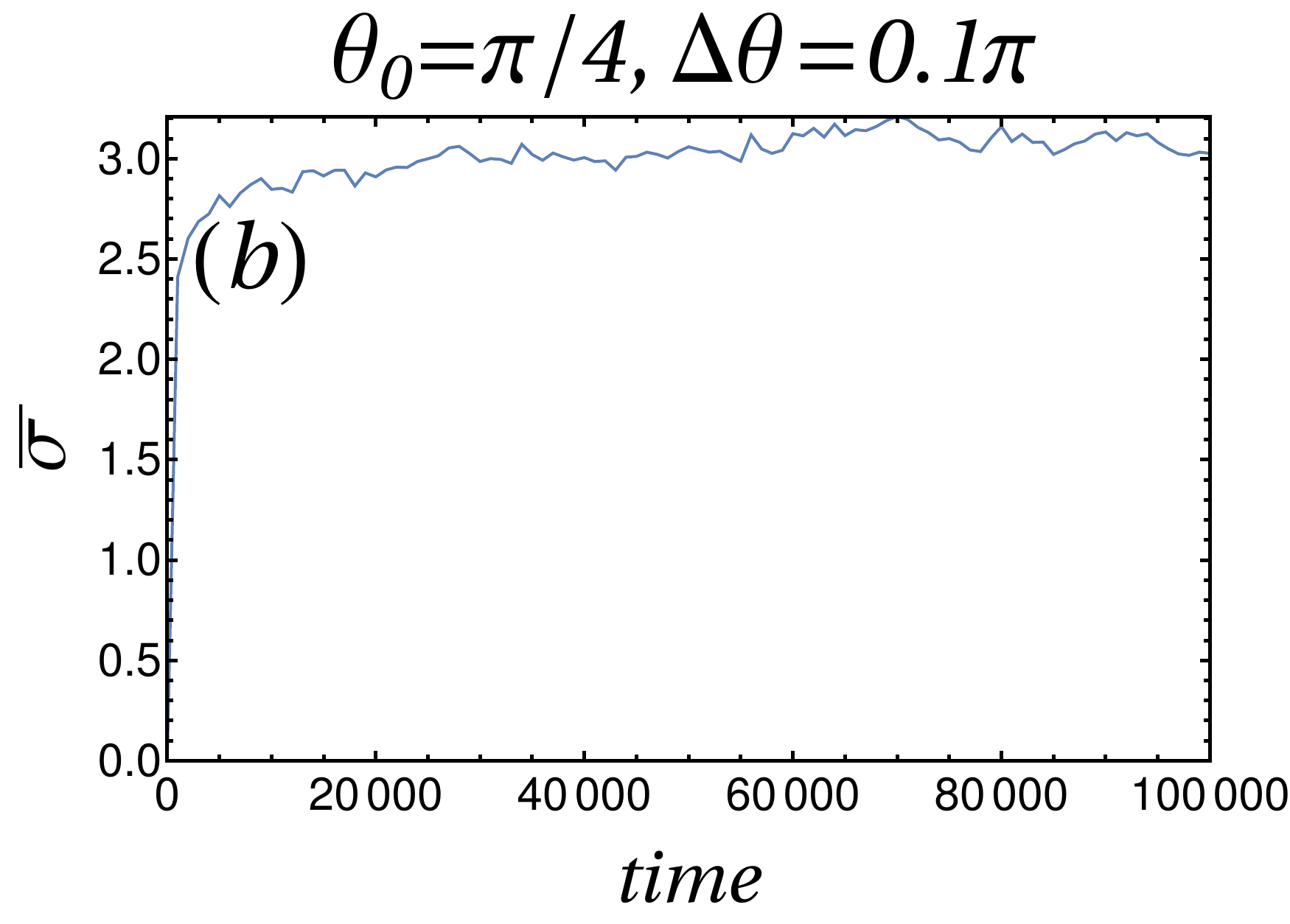}
\end{tabular}
   
  \caption{Dynamical properties of the disordered rim DC.  For rim sites, the parameter $\theta$ is chosen in a box centered around $\theta_0=\pi/4$ with width $\Delta\theta=0.1\pi$ and $\omega=\varphi=\beta=0$. For hub sites, we take a Grover coin at the critical flux $f_c=1/2$. (a): Typical probability distribution after $t=2000$ time steps for a state initially located on a hub site in the middle of the chain (note the semi-logarithmic scale). (b): Evolution of the standard deviation of the probability distribution averaged over $100$ realizations of the disorder.}
    \label{fig:sparim}
\end{figure}

In order to characterize the spreading of the walker, we study the standard deviation $\bar{\sigma}(t)$ of its wave function averaged over the disorder as a function of time, where $\sigma=\sqrt{\langle x^2\rangle-\langle x\rangle^2}$, $\langle. ..\rangle$ denotes the quantum average and $\overline{\cdots}$ the average over the disorder. The long-time behavior allows us to extract an exponent $\gamma$: $\overline{\sigma}\sim_{t\to\infty} t^\gamma$.

The asymmetry created between the path going through site $b$ or $c$ breaks the cages: in Figure~\ref{fig:sparim}-b, the standard deviation increases linearly at short times compared to its AB cage value. But at larger times, it saturates meaning that $\gamma=0$. The corresponding wavefunction shown in Figure~\ref{fig:sparim}-a features an exponential decay. The first effect of the disorder is to delocalize the walker. The second effect is to localize it. Anderson localization is expected (and was observed) for standard 1D QW with static disorder~\cite{Joye2010,Schreiber2011,Vakulchyk2017}. Here, we observe it for an AB-caged QW. A key difference is that, at the critical flux, the disorder-free model we consider is already localized with eigenstates having finite support.


\par
\begin{figure}[h]
\begin{tabular}{cc}
\multicolumn{2}{c}{\includegraphics[width=0.3\textwidth]{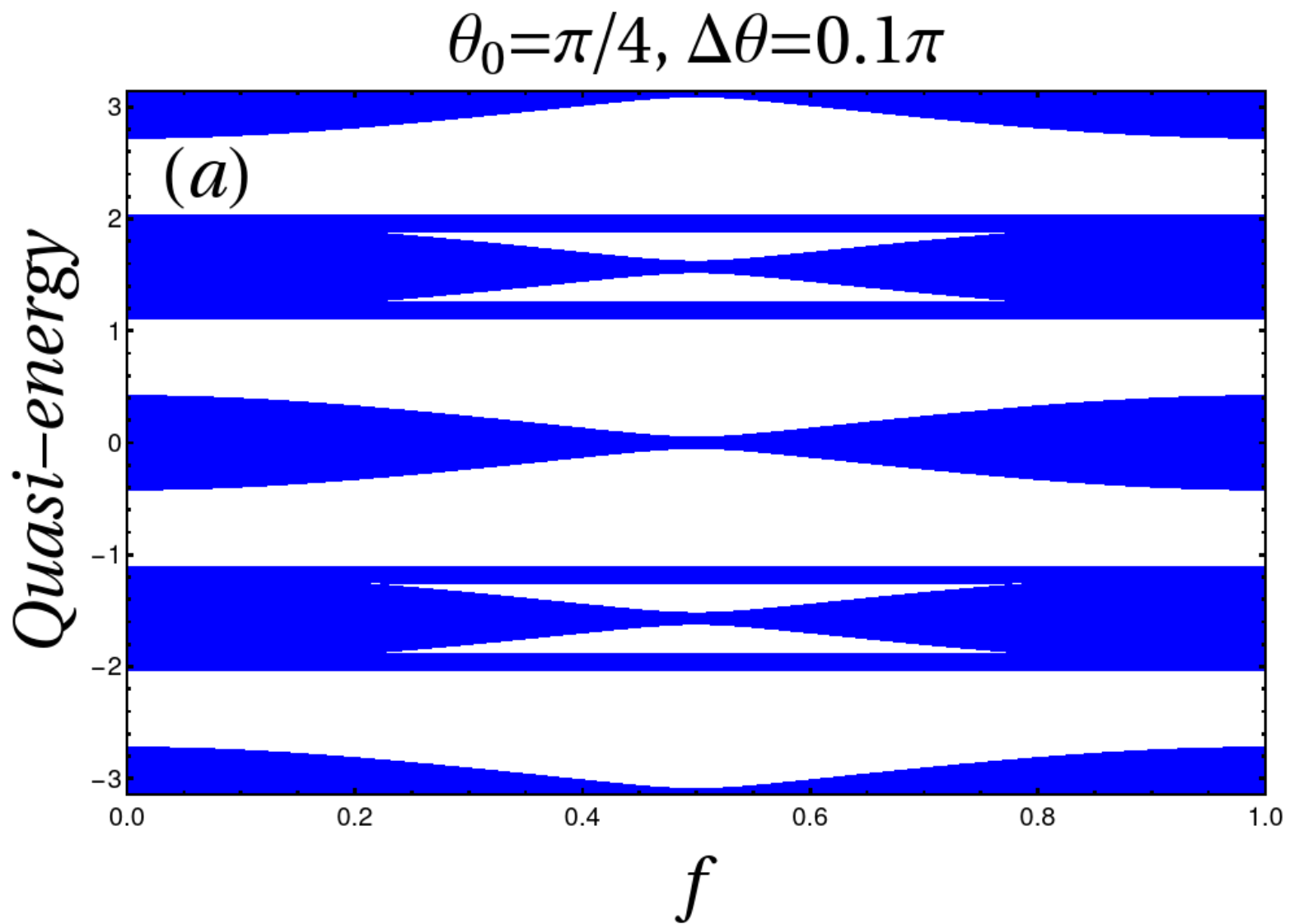}}\\
 \includegraphics[width=0.25\textwidth]{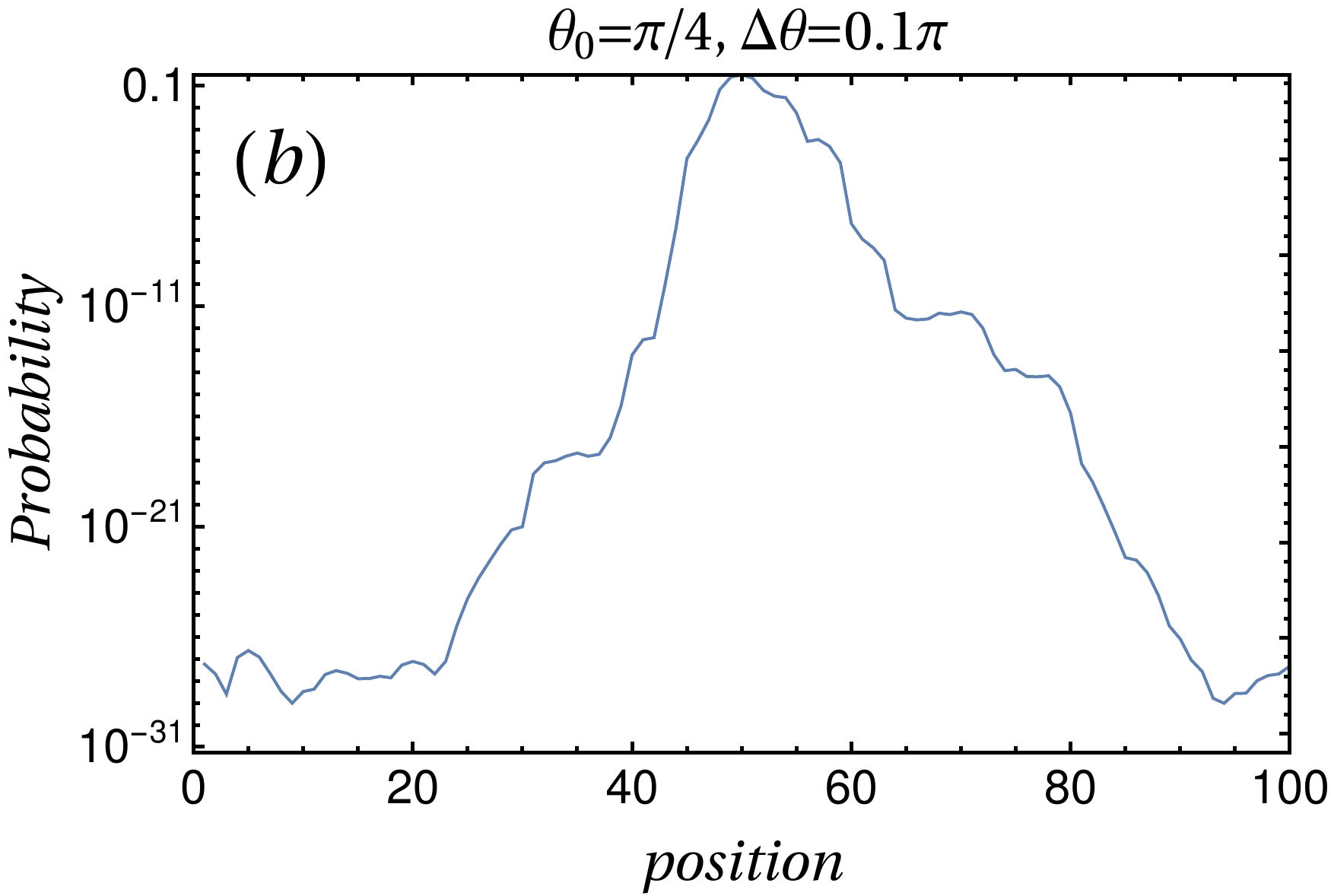}   & \includegraphics[width=0.25\textwidth]{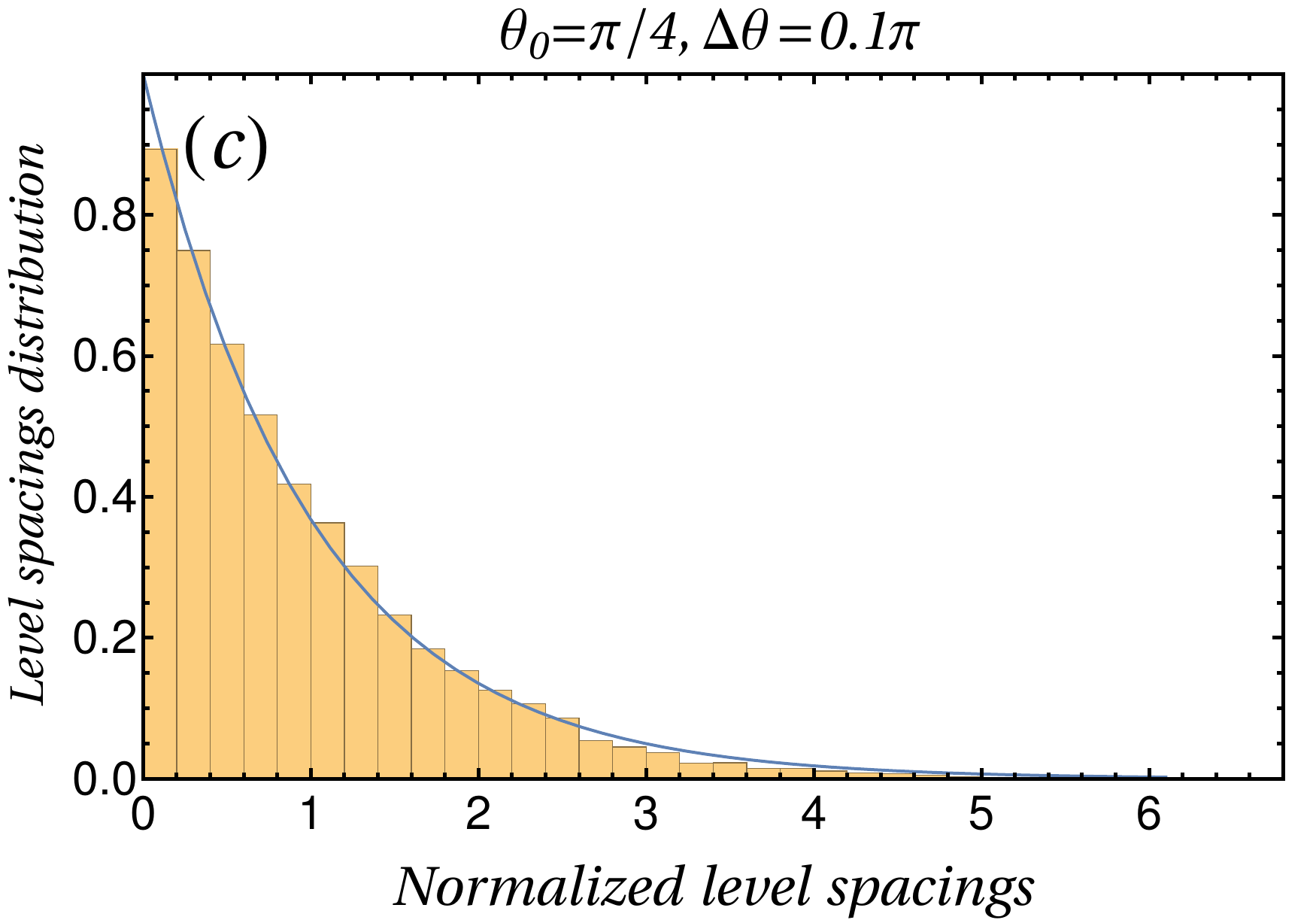}
\end{tabular}
   
  \caption{Spectral properties of the disordered rim DC. (a) Spectrum with respect to the reduced flux $f$. (b) A typical eigenvector (at the critical flux) shown as the probability to be on a cell in semi-logarithmic scale. (c) Level-spacing statistics (at the critical flux) averaged over $100$ disorder realisations for a system of size $L=100$. The blue curve is the Poisson distribution $P(s)=e^{-s}$.}
    \label{fig:spectresparim}
\end{figure}

 The numerically computed quasi-energy spectrum is shown in Figure~\ref{fig:spectresparim}-a. We see that the flat bands at the critical flux become of finite width because of disorder broadening. This is accompanied by exponential localization of the eigenvectors (see Figure~\ref{fig:spectresparim}-b) and Poisson-like distribution of the level differences (see Figure~\ref{fig:spectresparim}-c). All these features are consistent with Anderson localization of the walker. Below, we study the inverse participation ratio (IPR) in order to show that it is nevertheless different from standard Anderson localization.

To further analyze the effect of disorder on the model, we follow \cite{Vidal2001}, which studied the hamiltonian version of the AB cages on the DC. Disorder breaks the cages by coupling them. To quantify the degree of localization of the eigenstates, we use the IPR $I_2$. For normalized eigenfunctions $\ket{\Psi}$, it is defined by:
\begin{equation}
    I_2=\sum_{i\in\text{sites}}\left(\sum_{j\in\text{internal states}}|\bra{i,j}\ket{\Psi}|^2\right)^2
\end{equation}
\par When $I_2=1$, it means that the wave function is localized on one site. When it is equally distributed over the $N$ sites, $I_2=1/N^2$. The IPR is thus maximal for a localized state and minimal for a completely delocalized state.
\begin{figure}
    \centering
    \begin{tabular}{cc}
    \includegraphics[width=0.52\linewidth]{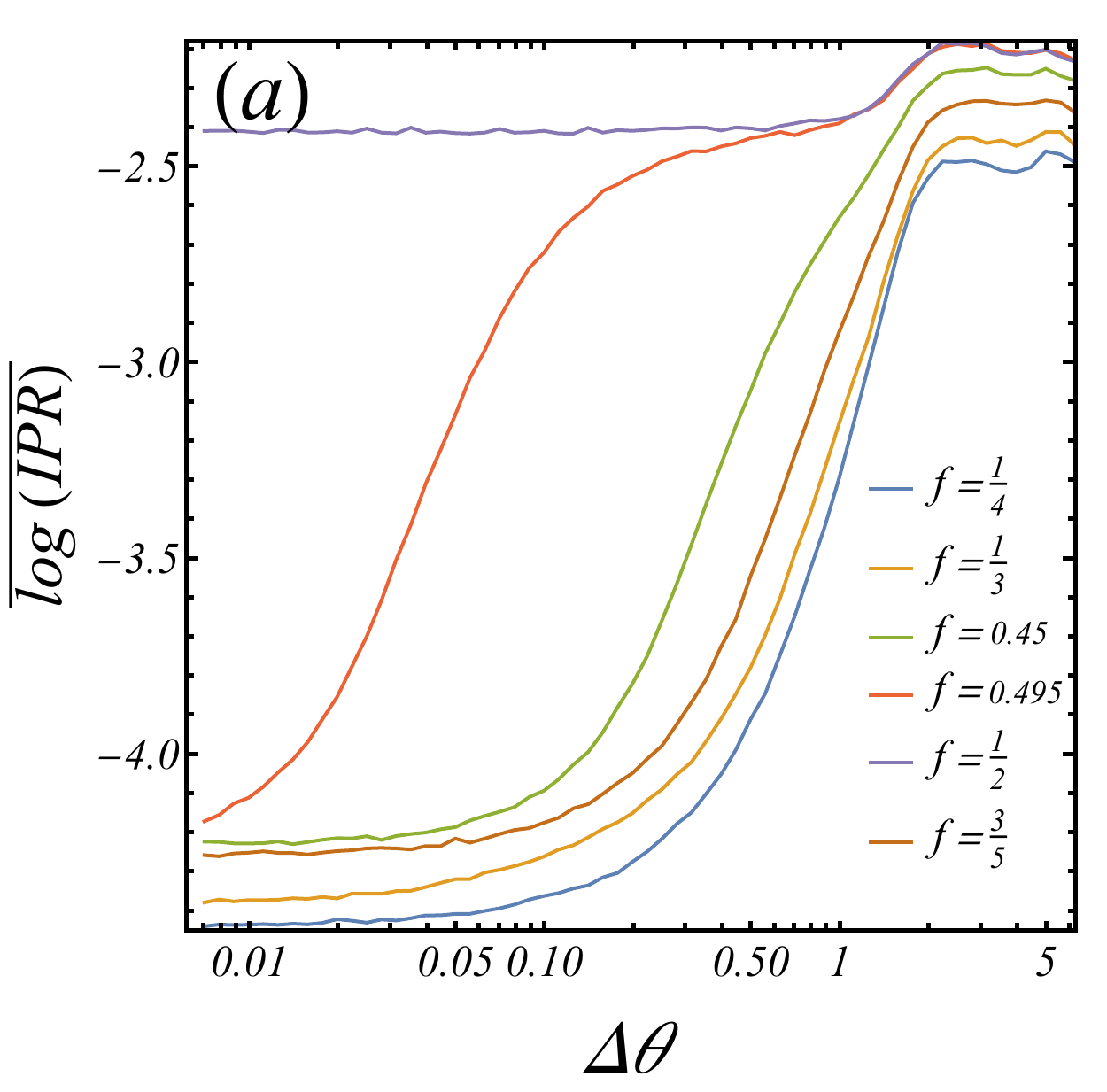}  \includegraphics[width=0.5\linewidth]{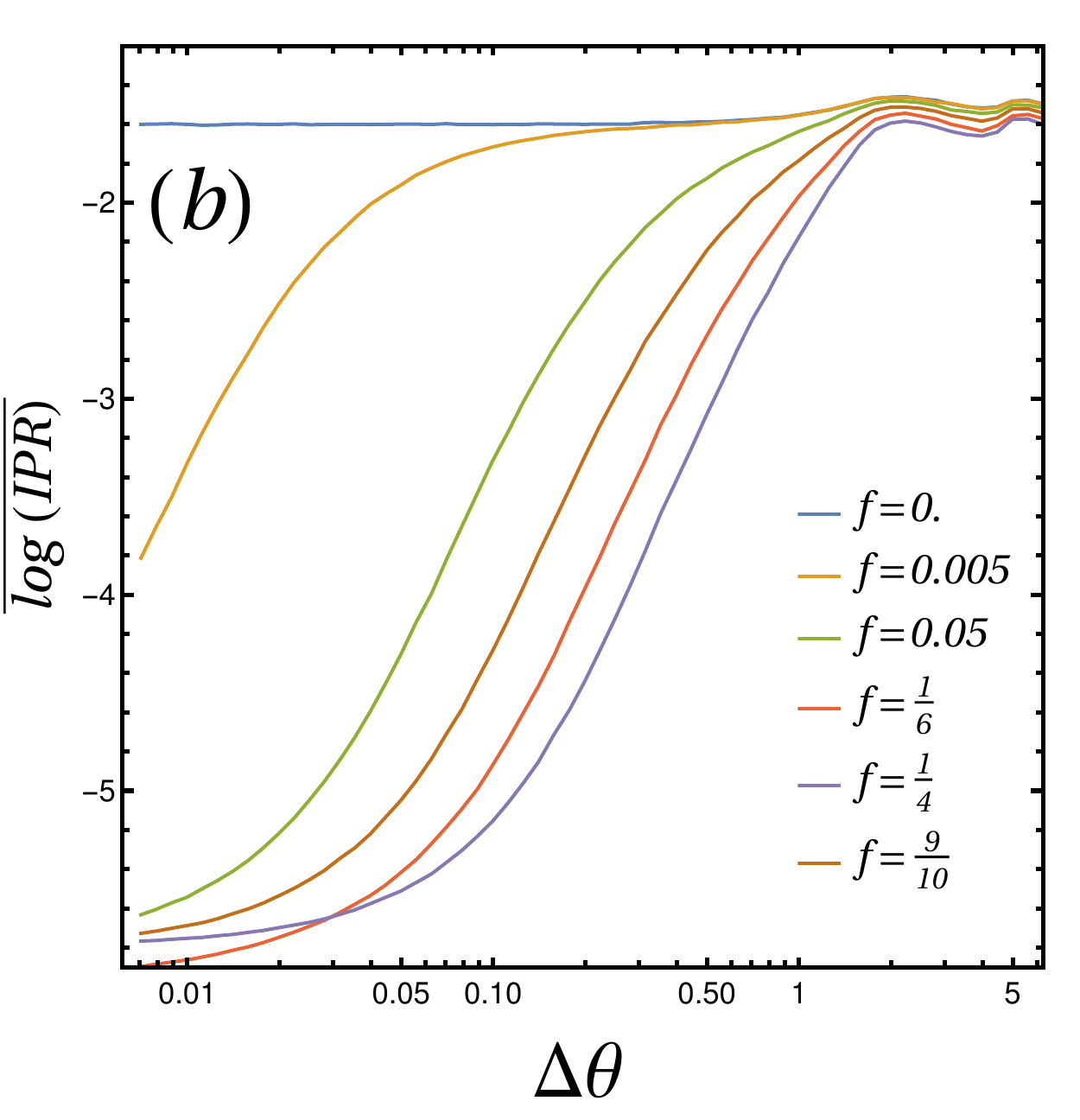}  \\

\end{tabular}
    \caption{IPR plotted versus the disorder strength $\Delta\theta$ at several fluxes $f$ for a DC with 600 sites. Log(IPR) is averaged over $100$ disorder realizations and over the quasi-energy. (a) Disorder only on the rim sites (Grover coin used on the hub sites) with $f=1/4,1/3,0.45,0.495,1/2,3/5$. (b) Disorder on both the rim and hub sites with $f=0,0.05,0.005,1/6,1/4,9/10$. The hub coin is defined in Eq.~\eqref{eq:haddes}.}
    \label{fig:iprdes}
\end{figure}

As done in Fig.~11 of~\cite{Vidal2001}, log(IPR) (averaged over both the eigenvectors and the disorder) is plotted as a function of the disorder strength $\Delta \theta$ in Fig.~\ref{fig:iprdes}. When the magnetic flux is tuned far away from the critical flux, we observe a monotonic growth of the IPR which corresponds to the usual Anderson localization as expected because the disorder-free model is a band model with ballistic propagation. However, at the critical flux we observe three different regimes (see Figure~\ref{fig:iprdes}-a). (i) When the disorder is weak, the IPR does not depend on $\Delta \theta$, meaning that the localization does not depend on the strength of the disorder. (ii) Then, for intermediate disorder, we observe a growth of the IPR corresponding to a localization of the eigenfunctions. (iii) Eventually, when $\Delta\theta>\pi$ (strong disorder), the IPR remains constant at its maximal value up to $\Delta\theta=2\pi$. 

For fluxes close to critical ($f=0.495$ in Figure~\ref{fig:iprdes}-a or $f=0.005$ in Figure~\ref{fig:iprdes}-b), we observe another intermediate regime where the IPR becomes equal to that at the critical flux. This is a regime where the width of the disorder-free band is of the same order as the strength of the disorder $\Delta\theta$. This is a reminiscent effect of AB caging away from the critical flux.

At the critical flux, the first two regimes (i) and (ii) are also observed in the hamiltonian version~\cite{Vidal2001}.
But the third regime (iii) is different. In the QW case, the disorder being bounded between $0$ and $2\pi$, the maximally disordered regime (in which the localization does no longer depend on the flux as in the hamiltonian system) cannot be reached. Such a regime occurs in the hamiltonian case because the localization is so strong that the particle is localised before encircling one loop and therefore does not experience the magnetic flux. In addition, in the QW case, the strength of disorder appearing only in cosine and sine function, its variation is bounded between $-1$ and $1$ and has reached these values already at $\Delta\theta=\pi$. Thus, for $\Delta\theta>\pi$, we do not expect the localization to vary anymore as observed on Figure~\ref{fig:iprdes}-a. 

In order to test the hypothesis that the disorder is too weak to reach the universal flux-independent regime of localization, we increase the effect of disorder by introducing extra disorder on hub sites. We use the following coin:
\begin{equation}
   H_2(\theta) \otimes H_2(\theta)= \begin{pmatrix}
    \cos\theta&\sin \theta\\
    \sin\theta&-\cos\theta\\
    \end{pmatrix}\otimes\begin{pmatrix}
    \cos\theta&\sin \theta\\
    \sin\theta&-\cos\theta\\
    \end{pmatrix} ,
    \label{eq:haddes}
\end{equation}
which gives back the $4\times4$ Hadamard matrix at $\theta=\pi/4$. The $\theta$ variable is, in the same way as on the rim sites, randomly chosen in a box distribution $[\theta_0-\Delta\theta/2,\theta_0+\Delta\theta/2]$ where $\theta_0=\pi/4$. At vanishing disorder, the critical flux is $f_c=0$ because $H_4$ coins are applied uniformly on the hub sites. We therefore look at the evolution of the IPR around this critical flux in Figure~\ref{fig:iprdes}-b. By disordering both the hub and rim sites, we realize that at high disorder, an almost common (i.e. flux-independent) saturating value is reached for the IPR. The maximal IPR is also larger than in the case without disorder on the hub sites, as there is a stronger localization of the eigenvectors since we introduced an additional disorder.

\section{Dynamical disorder}
\label{sec:dyndis}
\par So far, we have considered quenched disorder and its effect on the quantum spreading of the walker. We now turn to dynamical disorder. By pure dynamical disorder, we mean that at each time step $t$, a new set of spatially ordered coins is introduced that will trigger the quantum walk until the next time step $t+1$. This should clearly affect phase coherence phenomena based on several time steps evolution, like the caging effect at the critical flux.
Dynamically disordered coins for a quantum walker on a chain have been studied in~\cite{Ribeiro2004}. For random sequences of coin operations, they found a diffusive behaviour similar to a classical walk. The diffusive process has been observed for two types of disorder. The first one selects randomly one coin among two specific coins using an unbiased Bernoulli law. The second one uses a continuous variable parametrising the coin operator and randomly selected on a continuous set with variable width. In the DC case, we choose a dynamical disorder on hub sites to be similar to the first type of disorder by choosing randomly between $G_4$ and $H_4$ while the disorder on rim sites is chosen to be continuous.

\subsection{Dynamical disorder on hub sites}
\par Let us first consider disorder on the hub sites. As in the previous section on static disorder, we use a Bernoulli distribution of parameter $p_t$ to randomly select a coin between $H_4$ and $G_4$ at each time step. We thus have a probability $p_t$ to choose $G_4$ and $\bar{p_t}=1-p_t$ to choose $H_4$.

\par We keep track of the coin used on the hub sites (Grover or Hadamard) at each time. This forms a temporal chain of coin. In the same manner as done for static disorder, using the Table II of Appendix G in~\cite{Perrin2020}, one can infer rules on the temporal chain to determine the extension of the quantum walker on the diamond chain. They are relatively close to the one for the spatial disordered chain but we have to keep in mind that the conditions concern the time and not the positions:

\begin{itemize}
    \item First rule : The coin at time $t=0$ is irrelevant for the computation of the cage size.
    \item  Second rule : On the right-hand side, for $f=0$, the length of the QW after $T\geq2$ time steps is 2 + the number of Grover  coins between time $t=1$ and $t=T-1$. For $f=1/2$, we replace in the computation the number of Grover coins by the Hadamard ones.
     \item Third rule: On the left-hand side, for $f=0$, the length of the QW after $T\geq2$ time steps is 2 + the number of Grover coins between time $t=2$ and $t=T$. For $f=1/2$, we replace Grover coins by Hadamard coins.
\end{itemize}

\par From these rules it is possible to compute the probability that the walker is extended on $n$ sites at time $T$ at flux $f$. The minimal extension is $5$ sites, it corresponds to the temporal chain where only Hadamard coins appear (for the flux $f=0$) or Grover coins (for the flux $f=1/2$). When $n>5$, it means that Grover (resp. Hadamard) coin have been drawn for the flux $f=0$ (resp. $f=1/2$). 

\par Using the second and third rules at the flux $f=0$, we deduce that for each Grover coin located in the temporal chain between $t=2$ and $t=T-1$, the extension of the cage grows by one unit in both directions. If the Grover coin is located at $T=1$ (resp. $t=T$) it means that the extension of the cage is of one unit towards the right (resp. left). The same reasoning works for $f=1/2$ and the Hadamard coin. A computation similar to what was done in the spatial disorder section on hub sites leads us to an explicit expression for the probability of a cage of size $n$ at time $T$. For $n$ odd:

\begin{eqnarray}
    p_0(n,T)&=&\binom{T-2}{\frac{n-5}{2}}\bar{p_t}^{\frac{n-5}{2}}p_t^{T-\frac{n+3}{2}}\nonumber \\
    &+&\binom{T-2}{\frac{n-7}{2}}\bar{p_t}^{\frac{n-3}{2}}p_t^{T-\frac{n-3}{2}}\nonumber\\
     p_{1/2}(n,T)&=&\binom{T-2}{\frac{n-5}{2}}p_t^{\frac{n-5}{2}}\bar{p_t}^{T-\frac{n+3}{2}}\nonumber \\
     &+&\binom{T-2}{\frac{n-7}{2}}p_t^{\frac{n-3}{2}}\bar{p_t}^{T-\frac{n-3}{2}}
\end{eqnarray} and for $n$ even:
\begin{eqnarray}
    p_0(n,T)&=&2\binom{T-2}{n/2-3}\bar{p_t}^{n/2-3}p_t^{T-n/2+2}\nonumber\\
    p_{1/2}(n,T)&=&2\binom{T-2}{n/2-3}p_t^{n/2-3}\bar{p_t}^{T-n/2+2}
\end{eqnarray}

The average extension of the cages can then be computed with $\langle n(T)\rangle_f=\sum_n n \times p_f(n,T)$. When $T\to\infty$ the terms that mainly contribute to the sum are those around $n\sim T$. Thus, the average size of the walker's cages grows linearly with time: the cages are broken.

\par
\begin{figure}[h]
\begin{tabular}{cc}
 \includegraphics[width=0.25\textwidth]{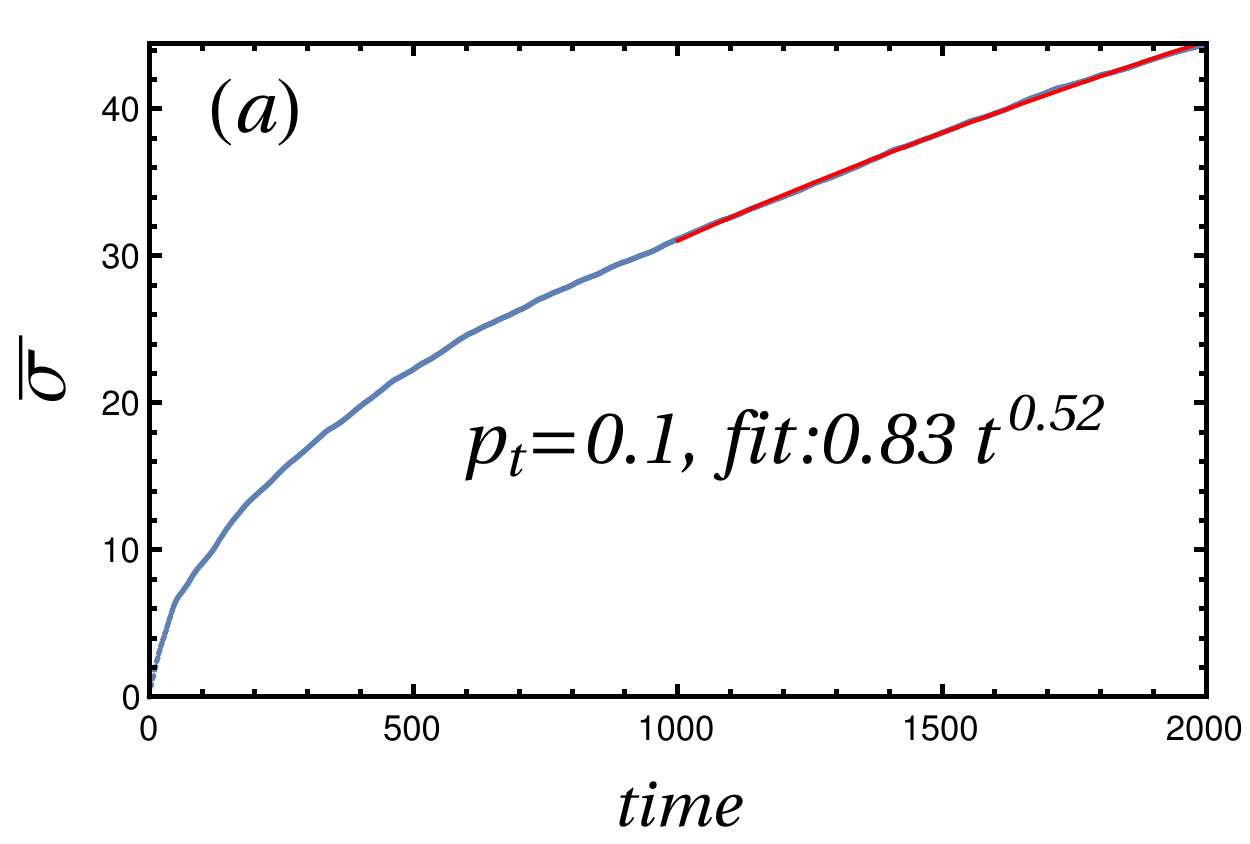}  & \includegraphics[width=0.25\textwidth]{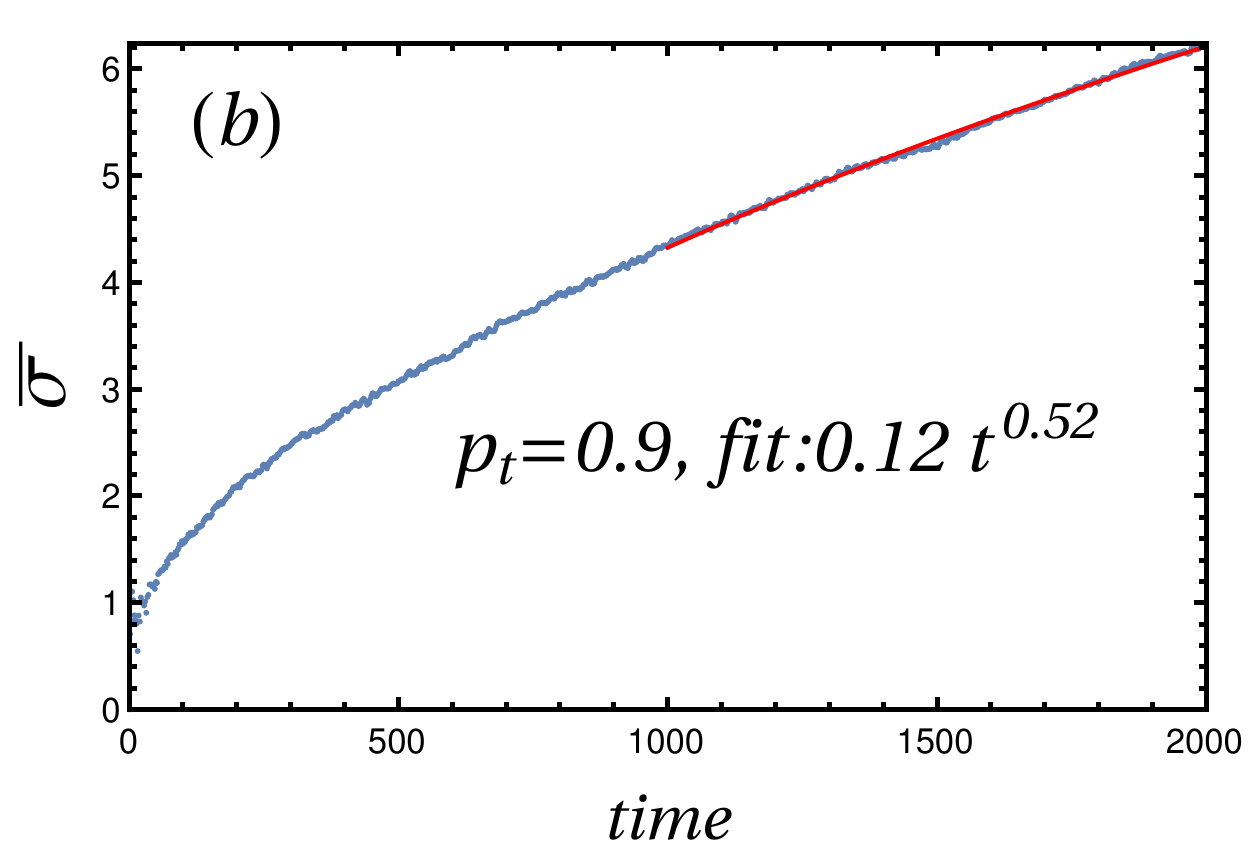}  
\end{tabular}
   
  \caption{Time evolution of the standard deviation averaged over 100 disorder realisations at flux $f=1/2$ for a dynamical disorder on the hub sites with (a) $p_t=0.1$ or (b) $p_t=0.9$. Rim coins have $\theta=\pi/4$ and $\varphi=\omega=\beta=0$. The fit (solid red curve) shows a diffusive exponent $\gamma\simeq 0.5$ and $p_t$ only affects the diffusion coefficient in front of the power law $t^\gamma$.}
    \label{fig:dynhub}
\end{figure}

 To characterize the dynamics of the wave function more precisely, we study its standard deviation as a function of time and we extract its power law. We obtain classical diffusion i.e. $\bar{\sigma}\underset{t\to\infty}{\simeq}{\sqrt{Dt}}$ (see Figure \ref{fig:dynhub}) as in the case of a quantum walk on a simple dynamically disordered chain. The parameter of the Bernoulli law controls the diffusion coefficient $D$: the more Grover coin there are (i.e. $p_t$ close to $1$ at $f=1/2$), the lower the diffusion coefficient. Dynamical disorder is expected to give back a classical behavior. The quantum coherences of the system vanish because the quantum evolution operator changes at every time steps. Note that there is no contradiction in the fact that $n\sim t$, which refers to the behavior of the cage extension (like a wave front), and that $\sigma \sim \sqrt{t}$, which refers to the way the wavefunction spreads.

\subsection{Dynamical disorder on rim sites}
\label{sec:dynrimdis}

\begin{figure}
\begin{tabular} {cc}
   \includegraphics[width=0.25\textwidth]{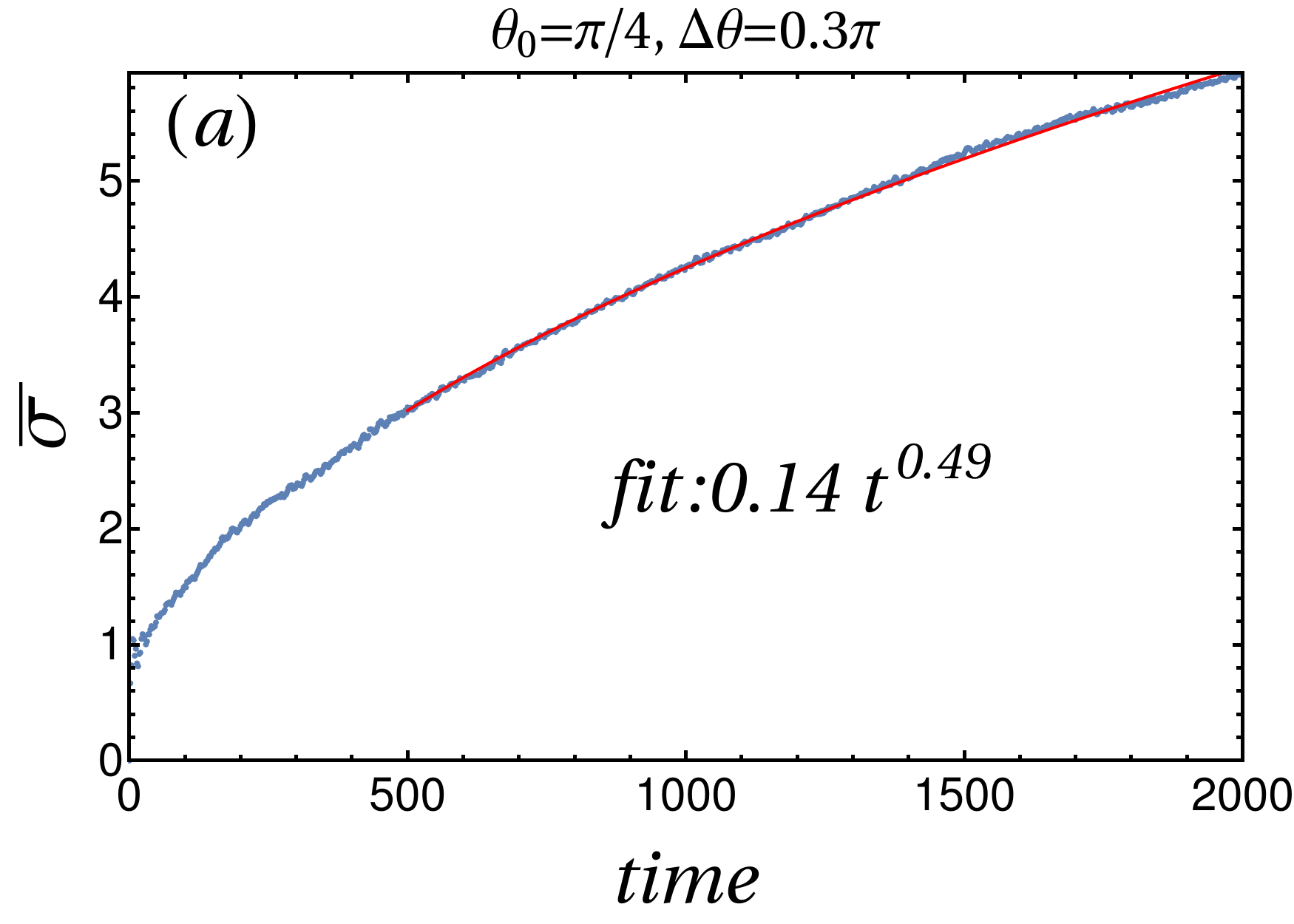}  &   \includegraphics[width=0.25\textwidth]{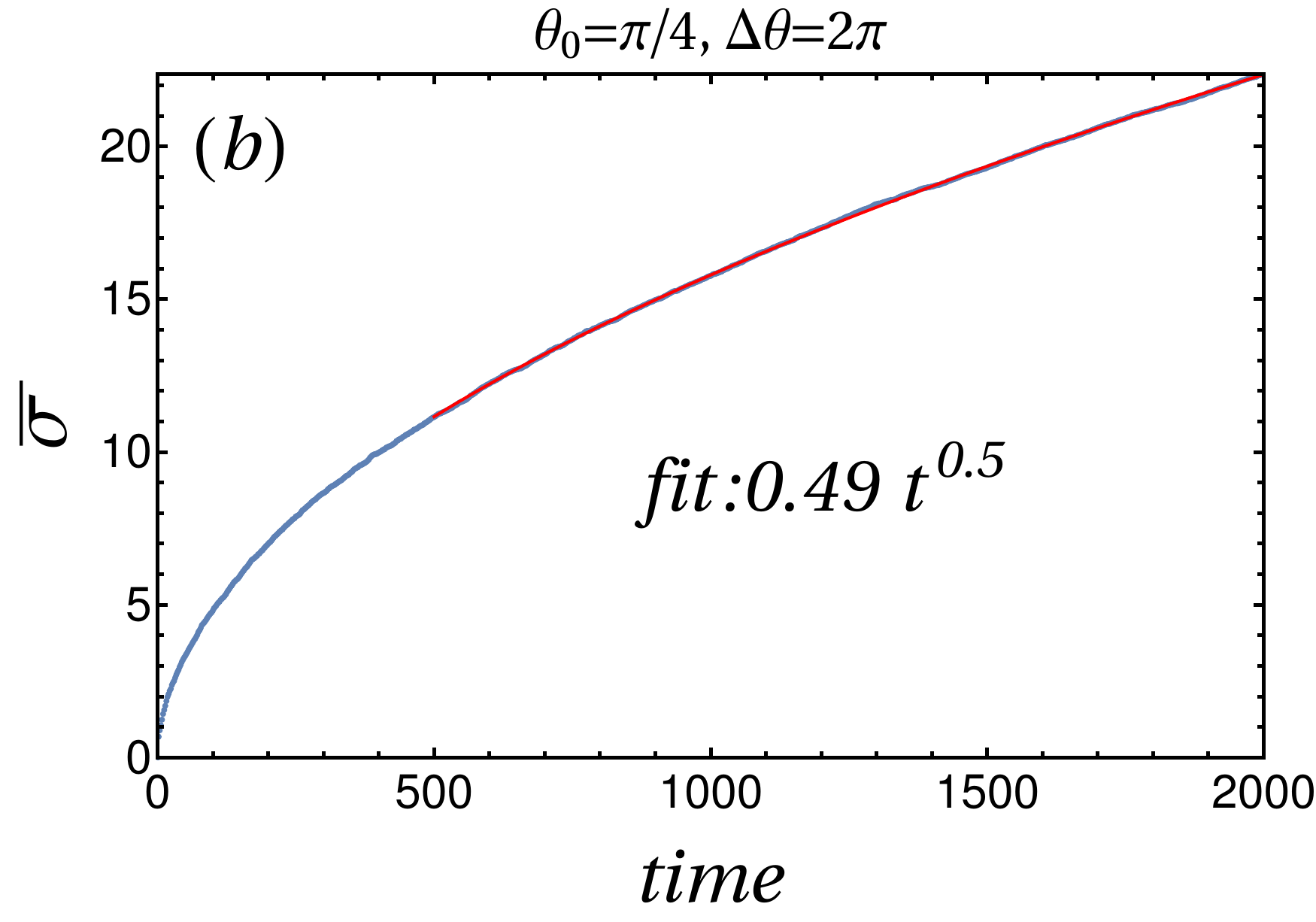}   \\
\end{tabular}

\caption{Time evolution of the standard deviation averaged over 100 disorder realisations at flux $f=1/2$ for a dynamical disorder on the rim sites. Parameters $\theta$ for $b$ and $c$ sites are chosen randomly and independently in the interval $\left[\theta_0-\Delta\theta/2,\theta_0+\Delta\theta_/2\right]$ at each time steps with $\theta_0=\pi/4$ (other parameters $\varphi=\beta=\omega=0$) and (a) $\Delta\theta=0.3\pi$ or (b) $\Delta\theta=2\pi$.}
\label{fig:dynrim}
\end{figure}

As in Sec.~\ref{sec:quenchedrimdis}, we choose to fix parameters $\beta=\varphi=\omega=0$ and consider only disorder on $\theta$. By unbalancing the coin on the $b$ and $c$ sites, the cage effect is destroyed. At each time step, two variables are randomly and independently chosen from the box-shaped distribution $[\theta_0-\Delta\theta/2,\theta_0+\Delta\theta/2]$. Each of these variables is applied uniformly to all rim sites of the same type ($b$ or $c$). The cages are thus broken and the number of sites visited by the walker increases linearly with time. We study the standard deviation of the probability distribution of the walker averaged over the disorder. As in the case of the dynamical disorder on the hub sites, we find a diffusive behavior $\bar{\sigma}\underset{t\to\infty}{\propto}t^{1/2}$ (see Figure \ref{fig:dynrim}).

The diffusive behavior can also be derived along the density matrix formalism $\rho(t)=\ket{\psi(t)}\bra{\psi(t)}$ where $\ket{\psi(t)}$ is the walker's wavefunction at time $t$~\cite{Ribeiroprivate}. The diagonal terms of the density matrix represent the probability to be on a site while the off-diagonal terms are called coherences, they quantify the degree of superposition of states.
When the dynamical disorder is turned on, the quantum walk operator is no longer time invariant. We expect the phase coherence will be lost and a classical behavior will be recovered, which is why we obtain a diffusive behavior as in the standard classical walk. For the density matrix, the off-diagonal terms tend to $0$ and it becomes essentially diagonal at long times. Details of the computation are given in Appendix~\ref{ap:diffusion}.

\section{Repeated measurements \label{sec:measure}}

\par A radically different way of introducing decoherence is by regularly measuring the position of the particle over time. In contrast to disorder, this does not modify the coin operation but directly perturb the wave function. One can choose to periodically measure its position, which implies a reduction of the wave packet, i.e. projecting its wave function onto an internal state of a site by following the associated probabilities. Even if the walker has a dynamic trapped in a cage of 5 cells between two measurements, the effect of the latter can send it to a site different from the initial one. Thus, it will be able to start a new trapped dynamics around this new site and reach sites it could not have accessed before. By repeating the process, the walker can then escape from its original cage. 

\begin{figure}
\begin{tabular}{cc}
      \includegraphics[width=0.53\linewidth]{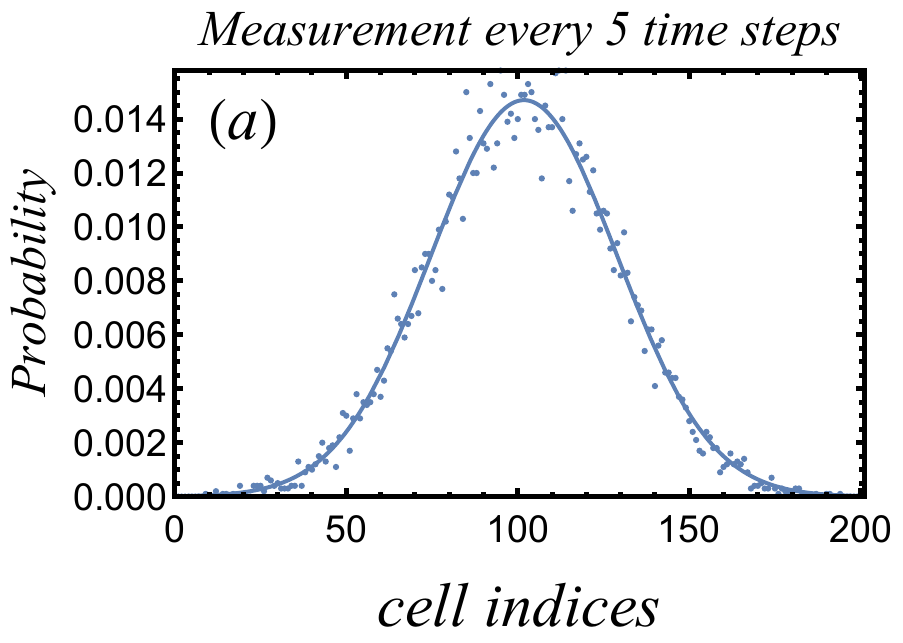} & \includegraphics[width=0.5\linewidth]{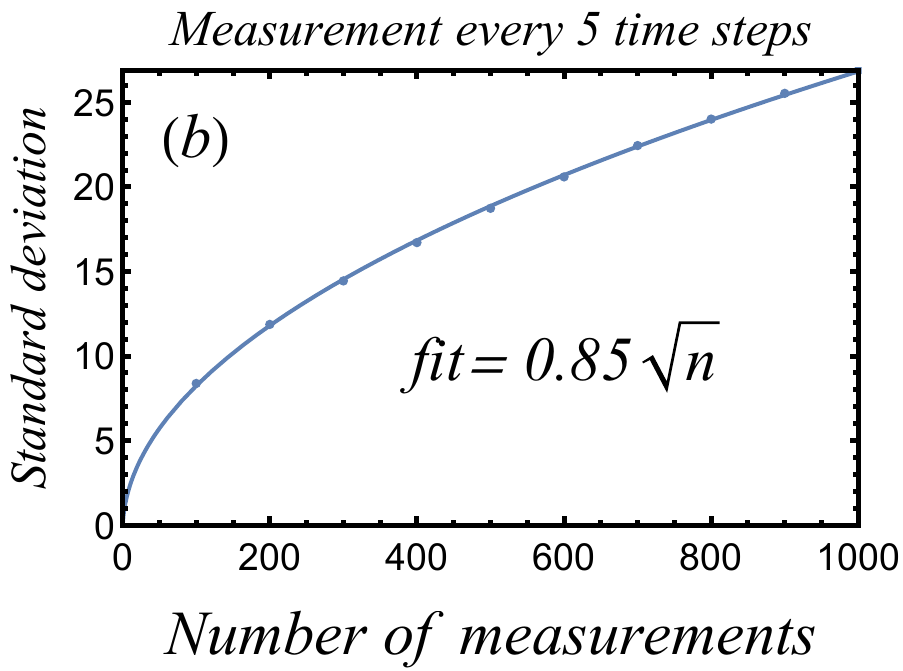} \\
\multicolumn{2}{c}{\includegraphics[width=0.53\linewidth]{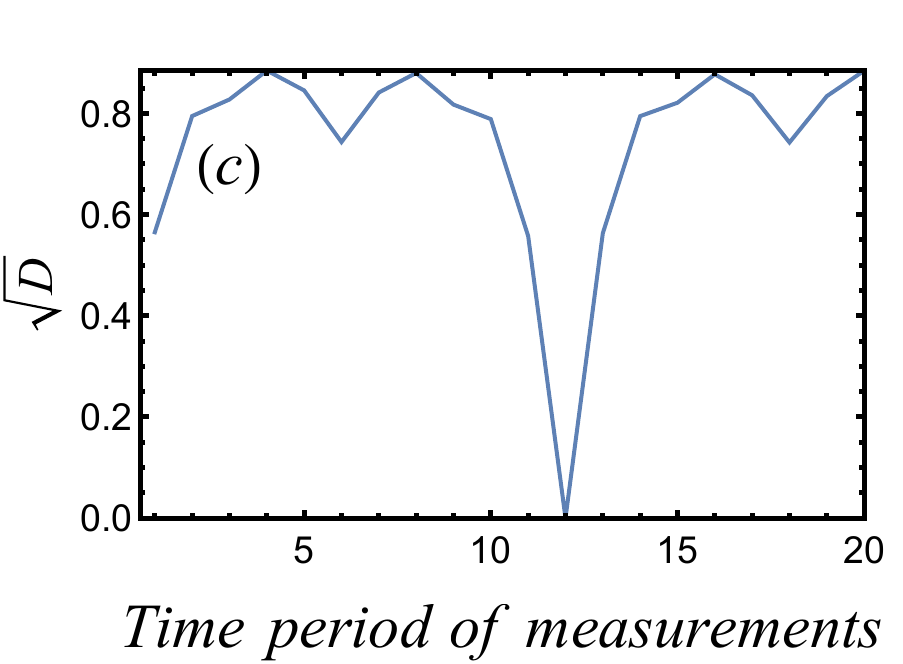}}
\end{tabular}
    \caption{Statistical properties of $10000$ different AB-caged QWs where the position is measured every 5 time steps. A Grover coin is used for hub sites  with $f=1/2$. Parameters of the rim coins are taken to be $0$ except for $\theta=\pi/3$. The initial spin configuration is $(\ket{R^+}+\ket{R^-})/\sqrt{2}$ and the position is on the cell of index $100$. (a): Distribution of the final positions after 1000 measurements. Points are numerical data and the solid line is a gaussian fit. (b): Standard deviation of the distribution of (a) with respect to the number of measurements. The solid line is a fit of the form $a\sqrt{n}$ where $n$ is the number of measurements and $a=\sqrt{D}=0.85$, where $D$ is a diffusion coefficient. (c): Evolution of $\sqrt{D}$ for different measurement periods $T$. The value $D=0$ for $T=12$ coincides with the period of the cage for $\theta=\pi/3$.}
    \label{fig:measure}
\end{figure}

\par Measurements reduce the wave packet and prevent superposition of states, so they are expected to cancel out quantum effects and give dynamics similar to classical systems. It should be pointed out, that unlike the other disorders introduced in the present article, this method gives rise to non-unitary dynamics because of the wave packet reduction. For the quantum walker on the diamond chain, we find a distribution close to the standard classical walker, a Gaussian centered around the initial site (see Figure \ref{fig:measure}-a) with a standard deviation growing proportionally to the square root of time or the number of measurements, see Figure \ref{fig:measure}-b). Figures \ref{fig:measure}-a,b are plotted for measurements every $5$ time steps but the result holds for any measurement period. There is a special case in which the cage has periodic dynamics, e.g., for a Grover coin on the hub sites at flux $f=1/2$ and parameters on the rim sites coin such as $\theta=\pi/3$ and $\varphi=\beta=\omega=0$, the period of the cages is $12$. When the period of the cages coincides with the period of the measurements, the diffusion coefficient vanishes (see Figure \ref{fig:measure}-c). The reason is that the measurement happens exactly when the probability that the walker is in its initial configuration is $1$. Except for this case, the quantum walker can escape from its cage diffusively thanks to the measurements.

\section{Dynamical and spatial disorder: subdiffusion \label{sec:subdiffusion}}

\par
\begin{figure}[!!h]
    \centering
    \includegraphics[width=0.5\textwidth]{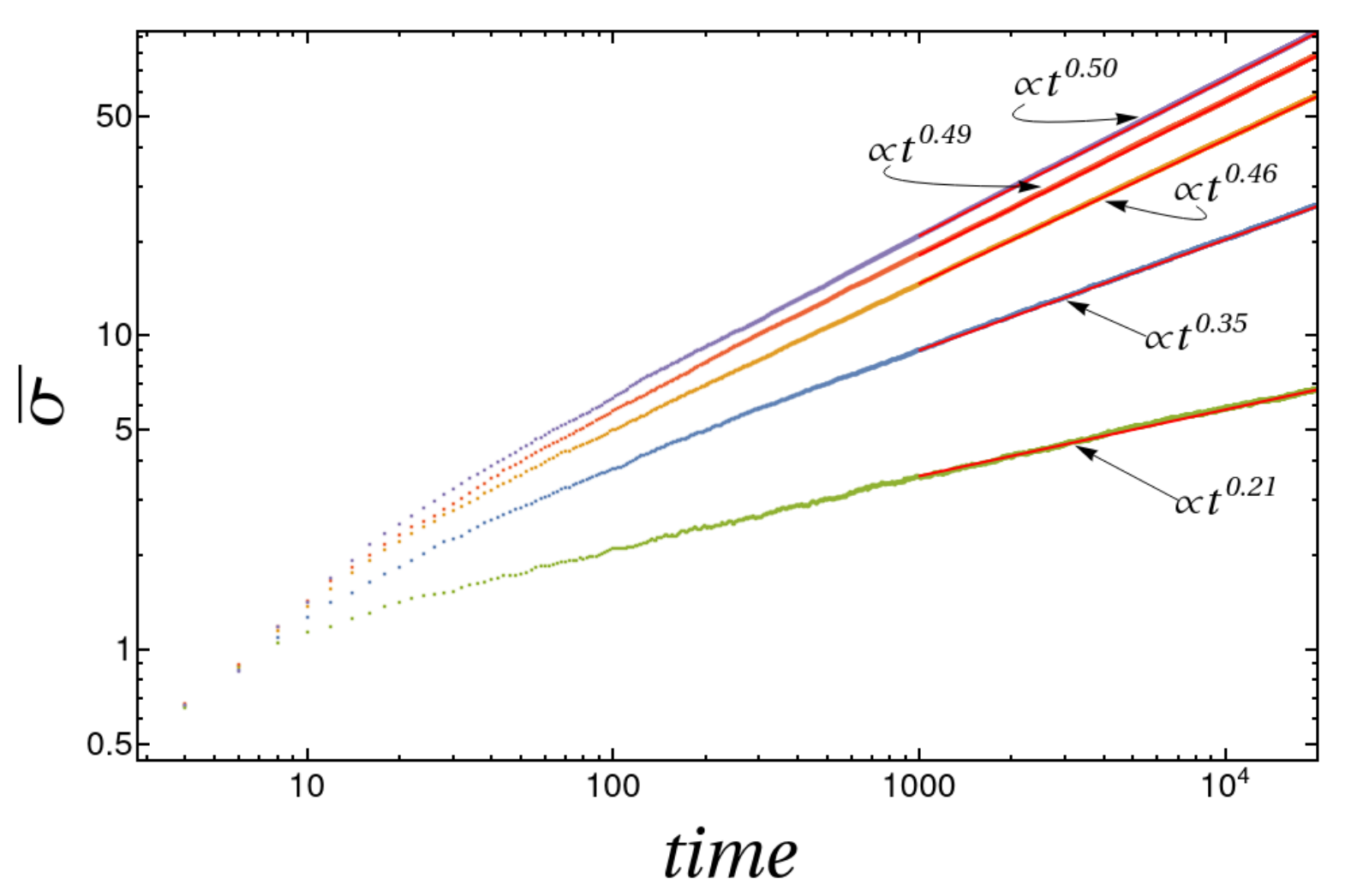}
    \caption{Standard deviations averaged over $100$ static disorder realizations of the variable $\theta_{as}$ taken in the distribution of Eq.~\eqref{eq:distanormale} with $\alpha=0.5,0,-0.5,-1,-2$ (from bottom to top curve) and a dynamical disorder on the variable $\theta_s$ taken in a simple box-shaped distribution. The double logarithmic scale highlights power laws. The fitted exponents $\gamma$ are, respectively, $0.21,0.35,0.46,0.49,0.50$ close to the predicted behaviors $t^{1/5}$, $t^{1/3}$, $t^{3/7}$, $\sqrt{t/\ln t}$, $t^{1/2}$.}
    \label{fig:sigma_anomalous}
\end{figure}

 In this section, we build a model leading to a more exotic long time behavior, namely subdiffusion with an exponent $0<\gamma<1/2$. Subdiffusion has already been obtained for one-dimensional quantum walks: either by using a static disorder that can be changed at each time step with a certain probability~\cite{Geraldi2021} or by introducing an additional degree of freedom, local spins, that interact with the quantum walker~\cite{Danaci2021}. Superdiffusion ($1/2<\gamma<1$), another kind of anomalous diffusion, has also been observed in one-dimensional QW by applying a quasi-periodic (Fibonacci) sequence of coin~\cite{Ribeiro2004}. Here, we propose another method to obtain anomalous diffusion by using a quantum walker on a caging diamond chain without any extra degree of freedom but a dynamical disorder changing at each time step and an extra quenched disorder (see Fig.~\ref{fig:sigma_anomalous}). We will see in the following that the quantum walk is mapped onto a classical walk presenting subdiffusion.

\par The classical walk in question is known as the symmetric random barrier model studied in~\cite{Alexander1981} (for a review of anomalous diffusion with classical random walks, see~\cite{Bouchaud1990}). The idea is that the transition rate (or jump probability) $W_{ij}$ between two neighboring sites $i$ and $j$ is randomly chosen in a certain distribution $\mathcal{P}(W)$. The transition rate simulates a potential barrier, the higher it is, the smaller is the potential barrier to cross. The behavior of the classical walk is determined by transition rates $\mathcal{P}(W)$ close to $0$ since it will be at these locations that the walker will spend most  of its time trying to cross barriers. Therefore, the asymptotic behavior of the distribution of transition rates near $0$ plays an essential role. In~\cite{Alexander1981}, the authors show that if $\overline{1/W}$ is finite i.e. $\mathcal{P}(W)\underset{W\to0}{\propto}W^{\mu-1}$ with $\mu>1$, one obtains a diffusive process. We recall that $\overline{\cdots}$ denotes the average over the disorder. On the contrary, when $\overline{1/W}$ is infinite i.e. $0<\mu<1$, there is an anomalous diffusion with an exponent $\gamma=\mu/(1+\mu)$. The marginal case $\mu=1$ leads to a behavior of the standard deviation in $\sqrt{t/\text{ln}t}$, which at very large times is hard to distinguish from a diffusive process. 

\par By combining a quenched disorder with a dynamical disorder on the rim sites in a specific way, the QW reproduces the classical symmetric random barrier model. Dynamical disorder maps the quantum walk onto a classical walk by killing coherences (as discussed in the previous section), while quenched disorder mimics the random transition rates of the classical model. On each rim site of the diamond chain, $\theta_b$ and $\theta_c$ are composed of a static and dynamical disordered variable such that $\theta_b(t,i)=\theta_s(t)+\theta_{as}(i)$ and $\theta_c(t,i)=\theta_s(t)-\theta_{as}(i)$. The probability for the quantum walker to escape from its cage is related to the bias of the variable $\theta$ between $b$ and $c$ sites of the same cell, $\theta_{as}(i)=(\theta_b(i)- \theta_c(i))/2$. More precisely, we can show that it is proportional to $\sin^2\theta_{as}(i)$. This quantity thus plays the role of random transition rates and we randomly choose $\theta_{as}$ according to a distribution:
\begin{equation}
    \left\{
    \begin{array}{ll}
       \mathcal{P}(\theta_{as})=\frac{1-\alpha}{2}\left(\frac{2}{\pi}\right)^{1-\alpha}|\theta_{as}|^{-\alpha}  &  \text{if} -\pi/2\leq\theta_{as}\leq\pi/2 \\
      \mathcal{P}(\theta_{as})=0   &\text{else} 
    \end{array}\right.
    \label{eq:distanormale}
\end{equation}

\par The prefactor in front of $|\theta_{as}|^{-\alpha}$ normalizes the distribution. The exponent $\alpha$ controls the behavior of the distribution in the neighborhood of $0$. Here, we have restricted $\theta_{as}$ to values between $[-\pi/2,\pi/2]$ in order to avoid multiples of $\pi$ which would play an important role because of the periodicity of the sine function involved in the cage leakage probability. 

\par The symmetric combination $\theta_s(t)=(\theta_b(t)+\theta_c(t))/2$ is chosen randomly at each time step (dynamical disorder) in a box-shaped distribution $[-\pi/2,\pi/2]$ and applied uniformly over the whole chain. The distribution of the cage leakage probability $(L\equiv\sin^2\theta_{as})$ near $0$ is:
\begin{align}
    \mathcal{P}(\theta_{as})\text{d}\theta_{as}\underset{\theta_{as}\to0}{\propto}|\theta_{as}|^{-\alpha}\text{d}\theta_{as}
    &=\frac{\left(\arcsin\sqrt{L}\right)^{-\alpha}}{\sqrt{L}\sqrt{1-L}}\text{d}L\nonumber\\
    &\underset{L\to0}{\sim}L^{-\frac{\alpha+1}{2}}\text{d}L
\end{align}

\par Thus we can identify the exponent $\mu$ of the symmetric random barrier model with the exponent $\alpha$ of the $\theta_{as}$ distribution: $\mu=(1-\alpha)/2$. We obtain an anomalous diffusion for $-1<\alpha<1$ whose exponent behaves as $\gamma=(1-\alpha)/(3-\alpha)$. When $\alpha<-1$, we find the standard diffusion process $\gamma=1/2$ and for $\alpha=-1$ we fall on the marginal case $\sqrt{t/\text{ln} t}$. 

\par In Figure~\ref{fig:sigma_anomalous}, we plot the different standard deviations, averaged over $100$ realizations of the disorder, of the wave function of the quantum walker as a function of different values of the power $\alpha=1/2,0,-1/2,-1,-2$. We obtain anomalous exponents $\gamma$ consistent with what is predicted by the classical symmetric random barrier model. For $\alpha=0.5$, the predicted anormal exponent is $0.2$ and the fit gives $0.21$. For $\alpha=0$ and $-0.5$, the values of the exponents of the respective fits are slightly higher, $0.35$ and $0.46$, than what is predicted by the theory, respectively $1/3$ and $3/7\simeq0.43$. For the marginal case $\alpha=-1$ which is supposed to give a behavior in $\sqrt{t/\text{ln}t}$, here we observe a power law $0.49$. The analysis should be refined to find the predicted result. Eventually, the expected diffusive behavior for $\alpha=-2$ is found.

\par Finally, we summarize in Table~\ref{tab:recapdes}, the main properties of the different perturbations studied in the previous sections as well as their effects on the quantum walker's wavefunction.

\begin{table*}[ht]
\begin{tabular}{|g{2cm}||m{2.cm}|m{2cm}|m{1cm}|m{1cm}|m{2.5cm}|m{3cm}|}
\hline 
Perturbation & \multicolumn{2}{c|}{Static disorder} &\multicolumn{2}{c|}{Dynamical disorder}& Repeated measurements & Spatial and dynamical disorder \\
 \cline{1-5}\cline{7-7}
 \hline
 Affected \newline sites   & \multicolumn{1}{c|}{hub} &\multicolumn{1}{c|}{rim}&\multicolumn{1}{c|}{hub}&\multicolumn{1}{c|}{rim}&\multicolumn{1}{c|}{hub and rim}&\multicolumn{1}{c|}{rim}\\
 \hline
 Cage breaking?&\multicolumn{1}{c|}{no}&\multicolumn{1}{c|}{yes}&\multicolumn{1}{c|}{yes}&\multicolumn{1}{c|}{yes}&\multicolumn{1}{c|}{yes}&\multicolumn{1}{c|}{yes}\\
 \hline
\multicolumn{1}{|c||}{\cellcolor{gris}Exponent $\gamma$}&\multicolumn{1}{c|}{0}&\multicolumn{1}{c|}{0}&\multicolumn{3}{c|}{1/2}&\multicolumn{1}{c|}{$0\leq\gamma\leq 1/2$}\\
\hline
\multicolumn{1}{|c||}{\cellcolor{gris}Mechanism}&Cages of tunable sizes& Anderson localization &\multicolumn{3}{c|}{Classical random walk}& Symmetric random barrier \\
\hline
\end{tabular}
\caption{Summary of the different perturbations studied for the AB QW cages with their principal properties.}
\label{tab:recapdes}
\end{table*}

\section{A second interacting particle \label{sec:2ndparticle}}

\par In this section, we are interested in the effect of a second distinguishable QW interacting with the first on the diamond chain. We obtain results similar to the Hamiltonian case, in particular the creation of a bound state which can move along the whole chain but with a strictly null probability of being separated by more than $4$ cells.

\par We first define the total Hilbert space of the two walkers and describe the network on which they evolve. Next, we review the case without interaction. Then, we introduce the interaction via a modification of the coin by a phase and we analyze the resulting spectra. 

\subsection{Hilbert space}

\par For the 1-body QW, there are three sites per cell: one hub site $a$ with four neighbours (two sites $b$ and two sites $c$)  and two rim sites  ($b$ and $c$ sites) with two neighbours $a$ sites. Here, we will take as the unit cell a hub site and the two rim sites located on its right. Each site is dressed with internal states equal to their coordination number. There are therefore 8 internal states per cell (see Figure~\ref{fig:diamond chain}). For a chain of length $L$ the size of the single-particle Hilbert space is $8L$.\\

\par For the 2-body  QW, the Hilbert space is the tensor product of two 1-body problem. We consider two distinguishable particles. The lattice is therefore a 2D structure with $8\times 8=64$ states distributed onto $3\times 3=9$ sites per cell which are : $(a,a)$, $(a,b)$, $(a,c)$, $(b,a)$, $(b,b)$, $(b,c)$, $(c,a)$, $(c,b)$, $(c,c)$. The first (resp. second) letter represents the site on which is the first (resp. second) quantum walker. The only ``hub-hub'' site (i.e. the two quantum walkers are each on a hub site) is the $(a,a)$ site, it contains $16$ internal states. The ``rim-rim'' sites (i.e. the two quantum walkers are each on a rim site) are the sites $(b,b)$, $(b,c)$, $(c,b)$ and $(c,c)$, each of them contain $4$ internal states. The sites of type ``hub-rim'' (i.e. the first walker is on a hub site and the second on a rim site): $(a,b)$ and $(a,c)$ or of type ``rim-hub'' (i.e. the first walker is on a rim site and the second on a hub site): $(b,a)$ and $(c,a)$ contain $8$ internal states. For a chain of length $L$, the Hilbert space is of dimension $64L^2$.

\begin{figure}[h]
   \includegraphics[width=0.5\textwidth]{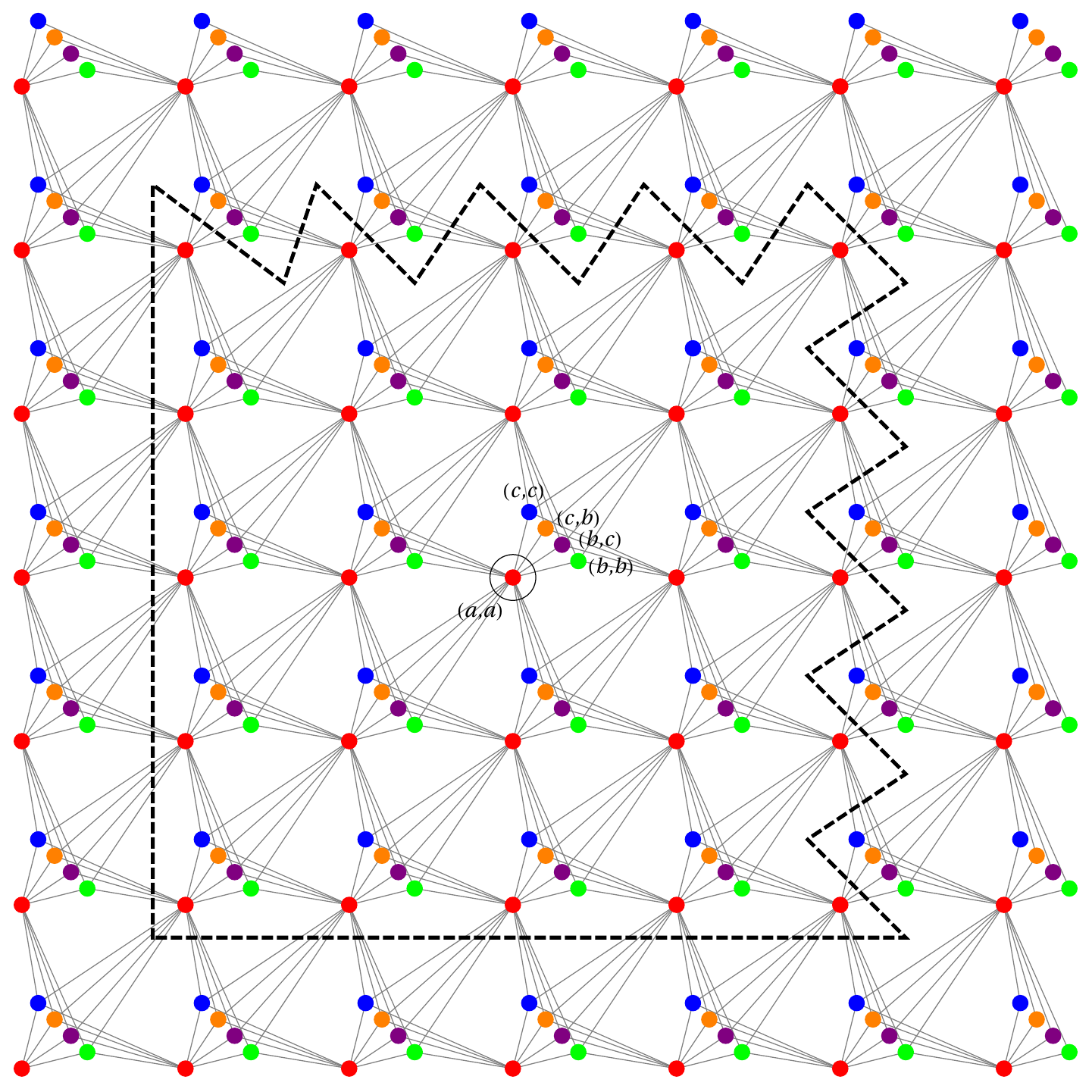}
  \caption{A piece of the first sublattice of the 2-body DC QW with sites of
type $(a,a)$, $(b,b)$, $(b,c)$, $(c,b)$ and $(c,c)$ (colored in red, green, purple, orange and blue). The horizontal (resp. vertical) direction represents the position of the first (resp. second) quantum walker. The black dashed
box indicates the maximal extension of a cage (at $f_c$, in the non-interacting case),
for an initial state localized at the $(a,a)$ site inside the circle.
The internal states are not represented.
}
    \label{fig:ssres1}
\end{figure}

\par The sites separate in two disconnected sublattices. The first one consists to start with both walkers on the same type of site (hub or rim sites), for instance, $(a,a)$. From this site, one can reach only rim-rim sites $(b,b)$, $(b,c)$, $(c,b)$ and $(c,c)$. From a hub site, a walker can either stay in its cell or go to the previous one. So the cell  from $(a,a)$ sites with $(n,m)$ cell indices is connected to cells $(n,m)$, $(n-1,m)$, $(n,m-1)$ and $(n-1,m-1)$. The previous rim-rim sites, namely, $(b,b)$, $(b,c)$, $(c,b)$ and $(c,c)$  with  cell index $(n,m)$ are only linked to $(a,a)$ sites with cell indices $(n,m)$, $(n+1,m)$, $(n,m+1)$ and $(n+1,m+1)$. Each cell is thus composed of $5$ sites: the hub-hub site and the $4$ rim-rim sites with a total of $16+4\times 4=32$ internal states in the cell. This first lattice is represented in Figure~\ref{fig:ssres1}. 

\par The second lattice is composed of the remaining sites $(a,b)$, $(a,c)$, $(b,a)$ and $(c,a)$. The construction is explicitly done in Appendix \ref{ap:lattice2} and represented Figure \ref{fig:ssres2}. 

\subsection{Non-interacting case}

\par Let $W_{2}=W_1\otimes W_1$ where $W_1$ is the unitary 1-body QW operator and $W_2$ is the unitary 2-body QW operator. For the 2-body QW, the shift operator is still hermitian and unitary and encodes the egdes of the sublattice 1 and 2 in Figure \ref{fig:ssres1} and \ref{fig:ssres2}. The coin operator is also the tensor product of 1-body coin. More precisely, on the 16-fold $(a,a)$ site it is $H_4\otimes H_4$ or $G_4\otimes G_4$. On the 4-fold $(x,y)$ sites where $x=b$ or $c$ and $y=b$ or $c$ it is $U_2(\theta,\varphi,\omega_x,\beta)\otimes U_2(\theta,\varphi,\omega_y,\beta)$.  On the 8-fold $(a,x)$ sites  the coin is $H_4\otimes U_2(\theta,\varphi,\omega_x,\beta)$ or $G_4\otimes U_2(\theta,\varphi,\omega_x,\beta)$. Eventually, the coin on the other 8-fold $(x,a)$ sites is $U_2(\theta,\varphi,\omega_x,\beta)\otimes H_4$ or $U_2(\theta,\varphi,\omega_x,\beta)\otimes G_4$.

\par
\begin{figure}[h]
\begin{tabular}{cc}
 \includegraphics[width=0.25\textwidth]{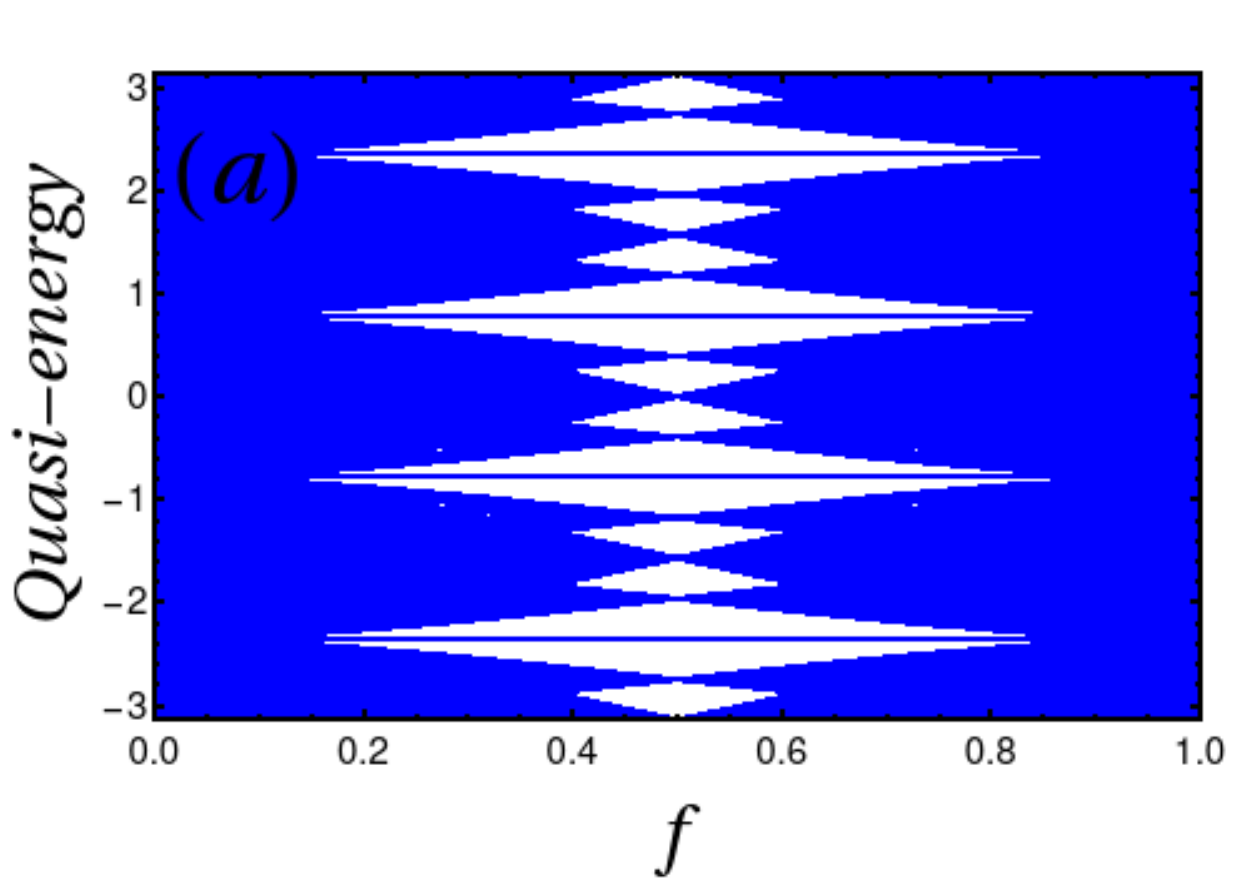}& \includegraphics[width=0.25\textwidth]{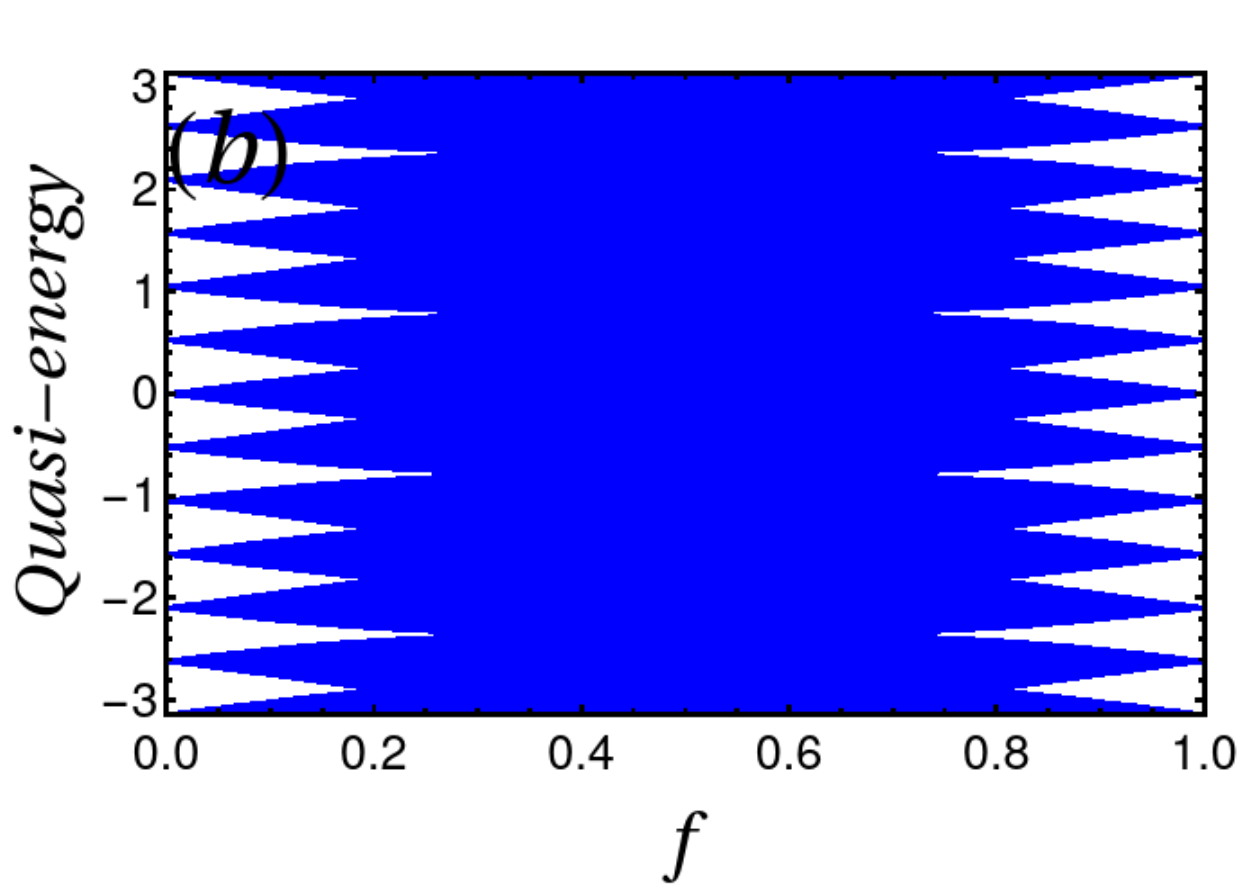}
\end{tabular}
  
  \caption{Spectra as a function of the magnetic flux $f$ for the two-body non interacting problem with (a) Grover or (b) Hadamard hub coins. Rim coins are $U_2(\pi/4,0,0,0)$.}
    \label{fig:papnonint}
\end{figure}

 Cages are expected to stand for the two non-interacting QW because the dynamics is the product of both 1-body dynamics  and the quasi-energies are the sum of both 1-body quasi-energies (see Figure \ref{fig:papnonint}): 
\begin{eqnarray}
W\ket{\psi}&=&W_1\ket{\psi_1}\otimes W_2\ket{\psi_2}\\
e^{-iE(k_1,k_2)}\ket{\psi}&=&e^{-i\varepsilon(k_1)}\ket{\psi_1}\otimes e^{-i\varepsilon(k_2)}\ket{\psi_2} \nonumber
\end{eqnarray}
where $\ket{\psi_1}$ and $\ket{\psi_2}$ are eigenvectors of the 1-body problem, $\ket{\psi}=\ket{\psi_1}\otimes\ket{\psi_2}$ are the 2-body eigenvectors of the non-interacting case with quasi-energies $E(k_1,k_2)=\varepsilon(k_1)+\varepsilon(k_2)$. Spectra of sublattice 1 and sublattice 2 are the same. Knowing the 1-body quasi-energies we can deduce the corresponding 2-body quasi-energies with their degeneracy. For instance, quasi-energies of the Grover QW at $f_c=1/2$ for $\theta=\pi/4$ and $\omega=\beta=\varphi=0$ is for the 1-body problem: $\varepsilon=0,\pm\pi/2,\pm3\pi/8,\pm5\pi/8,\pi$ degenerated only once. Quasi-energies of the 2-body problem are therefore $E=0,\pi$ 8-fold degenerated, $E=\pm \pi/2, \pm3\pi/8, \pm5\pi/8,\pm\pi/8,\pm7\pi/8$ 4-fold degenerated and $E=\pm\pi/4, \pm 3\pi/4$ 2-fold degenerated. On Figure \ref{fig:papnonint}-b,d representing spectra with respect to flux $f$ of the 2-body QW we still observe as expected a pinching of the energy levels at the critical flux $f_c=0$ (resp. $f_c=1/2$) for the Figure \ref{fig:papnonint}-b (resp. Figure \ref{fig:papnonint}-d)

\par At the critical flux $f_c$, the 2-body non-interacting eigenstates can be denoted by $\ket{n_1,\varepsilon_1,n_2,\varepsilon_2}$ where $\ket{n_i,\varepsilon_i}$ is the 1-body maximally confined eigenstate of the particle $i$ described in Sec.~\ref{sec:QWDC} and shown Figure~\ref{fig:eigenvectors}. We recall that this maximally confined eigenstate has weight on 3 cells of the diamond chain, $n_i$ is the index of the middle cell and $\varepsilon_i$ its associated quasi-energy. 

\subsection{Interacting case}
\label{sec:interact}
We add an on-site interaction to the 2-body QW. As a consequence only the dynamics on the first sublattice is affected by the interaction. The $(a,a)$, $(b,b)$ and $(c,c)$ sites localized on the diagonal ($n=m$) are the only sites perturbed by the interaction. The standard way to simulate an on-site potential for QW, is to multiply coins on these sites by a complex exponential $e^{i\phi}$~\cite{Ahlbrecht2011,Berry2011,Bisio2018,Verga2018,Verga2019,Toikka2020,Malishava2020} where $\phi$ 
mimics the on-site interaction $U$ in the Hubbard model.\\

\par We then focus on the first sublattice where the dimension of the associated Hilbert space is $32L^2$. In the non-interacting  case, we found $64L^2$ eigenvectors living on the two sublattices.  These eigenvectors projected on one of the two sublattices do not form a basis anymore because they are not linearly independent. It is then necessary to select $32L^2$ independent vectors among these $64L^2$.  A simple solution amounts to select only states corresponding to the quasi-energies $\varepsilon_1=0,\pi/2,3\pi/8,-3\pi/8$.\\

\par The translation invariances along  the natural direction $x$ (position of the first QW) and $y$ (position of the second QW) are broken due to the interaction. However, the translation along the $x_+=\frac{x+y}{2}$ direction remains a symmetry. One can index the problem then by the momentum $k_+$ associated to the $x_+$ direction. As a consequence for each $k_+$, the problem becomes 1D in the $x_-=x-y$ direction with a potential on the site $x_-=0$.
\par The total symmetry of the wavefunction under the exchange of both particles is conserved. We denote the symmetric $S$ and antisymmetric $AS$ part of the eigenvectors (without interaction):
\begin{equation}
    \ket{n_1,\varepsilon_1,n_2,\varepsilon_2}_{S/AS}=\frac{1}{\sqrt{2}}(\ket{n_1,\varepsilon_1,n_2,\varepsilon_2}\pm(\ket{n_2,\varepsilon_2,n_1,\varepsilon_1})
\end{equation} if $\varepsilon_1\neq\varepsilon_2$ or $n_1\neq n_2$. One can also express them in function of the $k_+$ representation:
\begin{equation}
\ket{\psi_j(k_+,\varepsilon_1,\varepsilon_2)}_{S/AS}=\frac{1}{\sqrt{L}}\sum_{n=0}^{L-1}\e^{-\im nk_+}\ket{n,\varepsilon_1,n+j,\varepsilon_2}_{S/AS}
\end{equation} where $j\in[0,L-1]$ is an integer and $n+j$ is defined modulo $N$.
\par  When the initial state is far from the coordinate $x_-=0$, cages of the two QW do not overlap, they remain unchanged. Actually, only few states are affected by the interaction: $\ket{\psi_0(k_+,\varepsilon_1,\varepsilon_2)}_{S/AS}$, $\ket{\psi_{1}(k_+,\varepsilon_1,\varepsilon_2)}_{S/AS}$ and $\ket{\psi_{2}(k_+,\varepsilon_1,\varepsilon_2)}_{S/AS}$ and they are not eigenvectors anymore. However subspaces generated by $\{\ket{\psi_0(k_+,\varepsilon,\varepsilon')}_{S/AS}\}_{\varepsilon,\varepsilon'}$,  $\{\ket{\psi_1(k_+,\varepsilon,\varepsilon')}_{S/AS}\}_{\varepsilon,\varepsilon'}$ and  $\{\ket{\psi_2(k_+,\varepsilon,\varepsilon')}_{S/AS}\}_{\varepsilon,\varepsilon'}$ (with $\varepsilon,\varepsilon'$ running through the different possible quasi-energy values) remain orthogonal to each other.

\par
\begin{figure}[h]
   \includegraphics[width=0.5\textwidth]{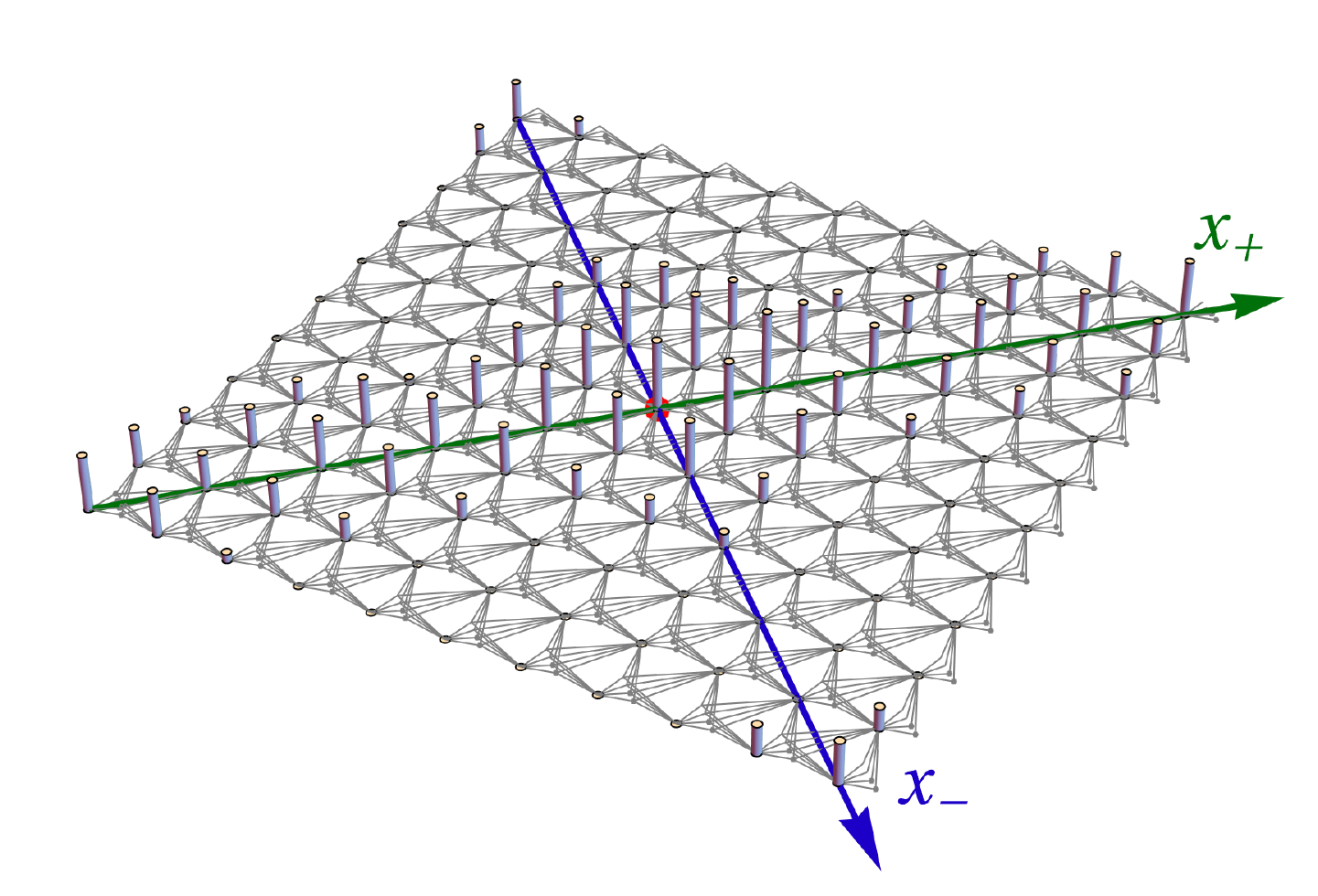}
  \caption{The presence probability on each site is shown after $T=50$ time steps using Hadamard and $U_2(\pi/4,0,0,0)$ coins at $f_c=0$ with an interaction $\phi=2$. The QW starts from the $(a,a)$ site at the cell of index (4,4).  The non-vanishing probability in the corner of the anti-diagonal is due to periodic boundary conditions. }
    \label{fig:animint}
\end{figure}

 In the $x_+$ direction, at $x_-=0$, the interaction completely changes the dynamics of the QW: the cage is destroyed as shown in Figure~\ref{fig:animint} and is replaced by a delocalized bound state. 

\par
\begin{figure}[h]
\begin{tabular}{cc}
 \includegraphics[width=0.25\textwidth]{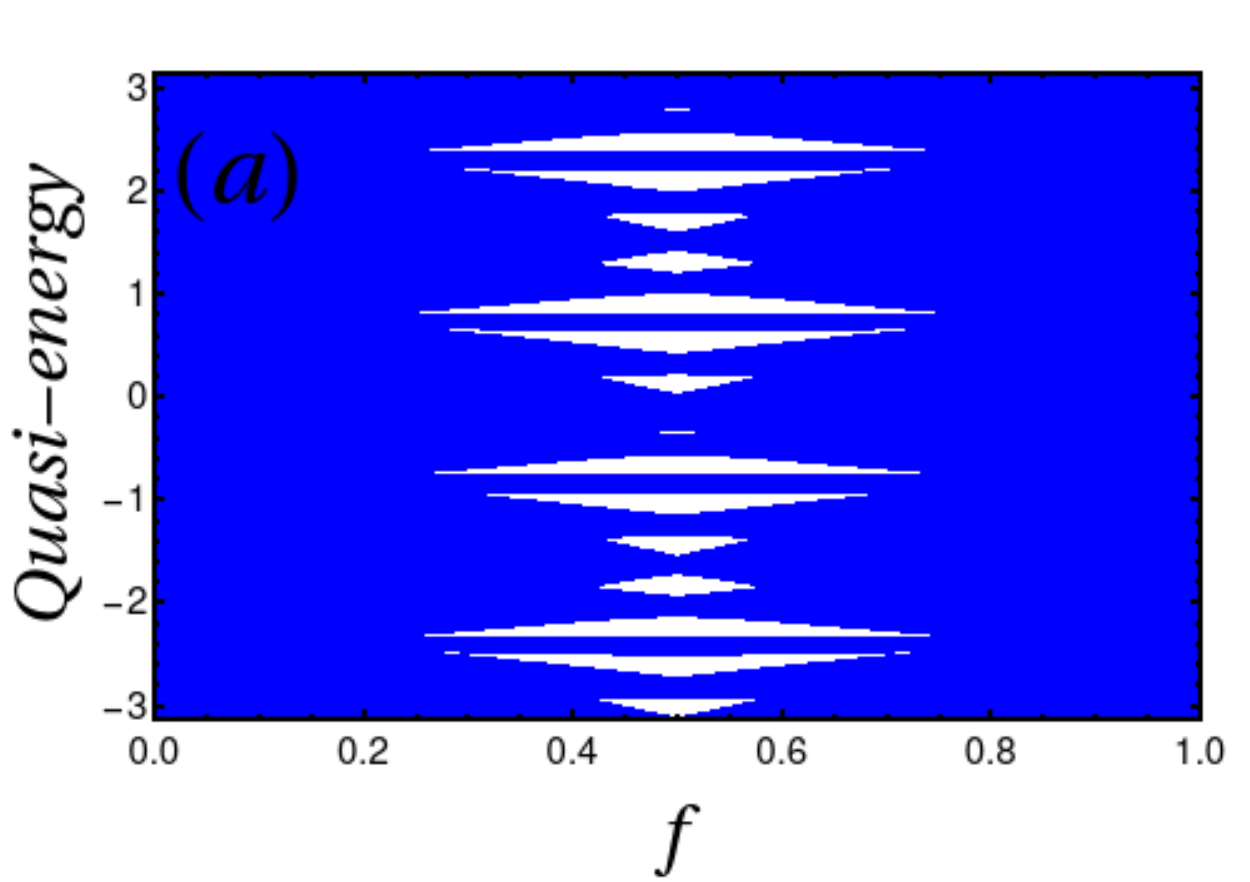}& \includegraphics[width=0.25\textwidth]{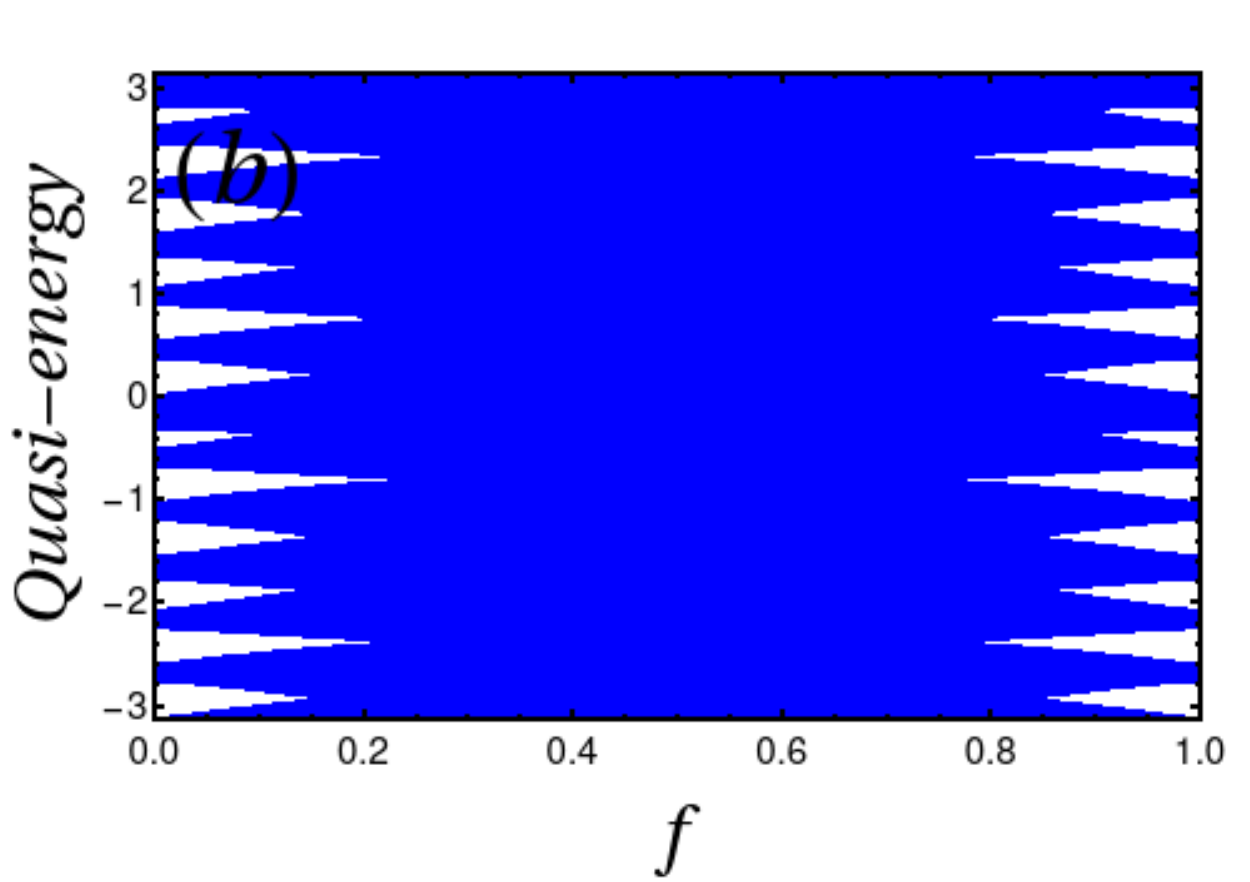}
\end{tabular}
  \caption{Spectra as a function of the flux $f$ for interacting QW using (a): a Grover coin, (b) a Hadamard coin, and a coin $U_2(\pi/4,0,0,0)$ on rim sites with an interaction $\phi=0.1\pi$. There is no pinching anymore at the critical flux if we compare to the non interacting case (see Figure~\ref{fig:papnonint}).}
    \label{fig:papint}
\end{figure}

 Figure~\ref{fig:animint} highlights the creation of an extreme bound state which can move into the whole system but where the two QW can not be separated more than four cells away. As a consequence of the interaction, the quasi-energy spectra of these QW does not pinch anymore at the critical flux as shown in Figure \ref{fig:papint}. There are no flat bands at the critical flux but dispersive ones  which enhance the destruction of cages in the $x_+$ direction. Quantum walkers leak out from the cage ballistically.\\

\par
\begin{figure}[h]
\begin{tabular}{cc}
 \includegraphics[width=0.25\textwidth]{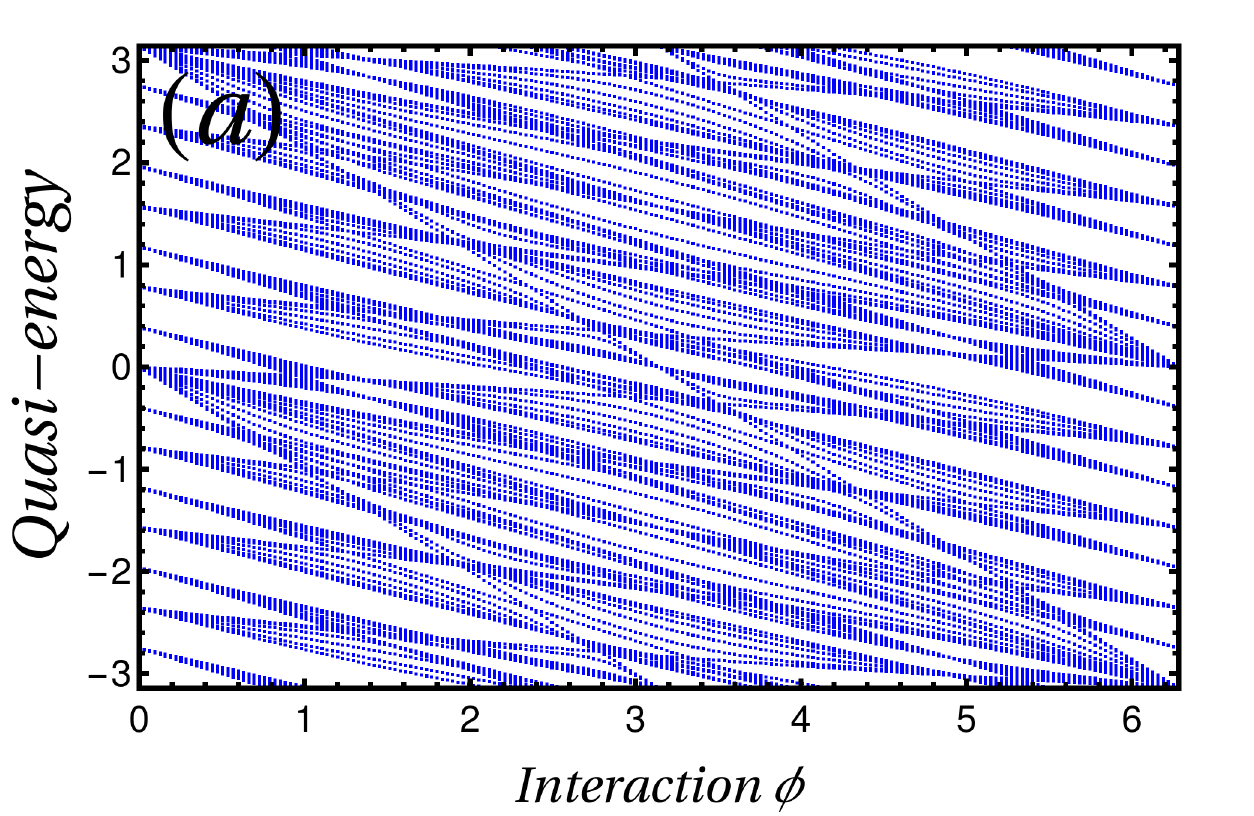}    &  \includegraphics[width=0.25\textwidth]{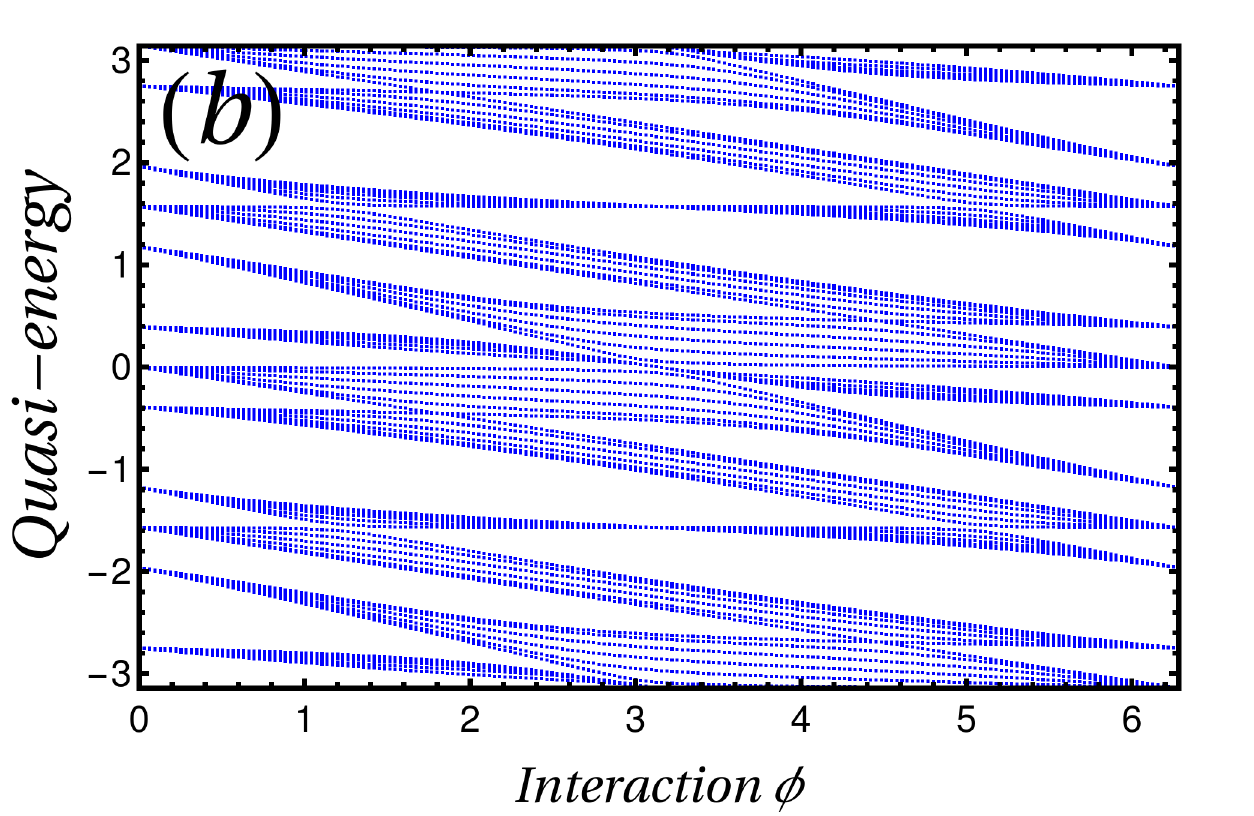} \\ \includegraphics[width=0.25\textwidth]{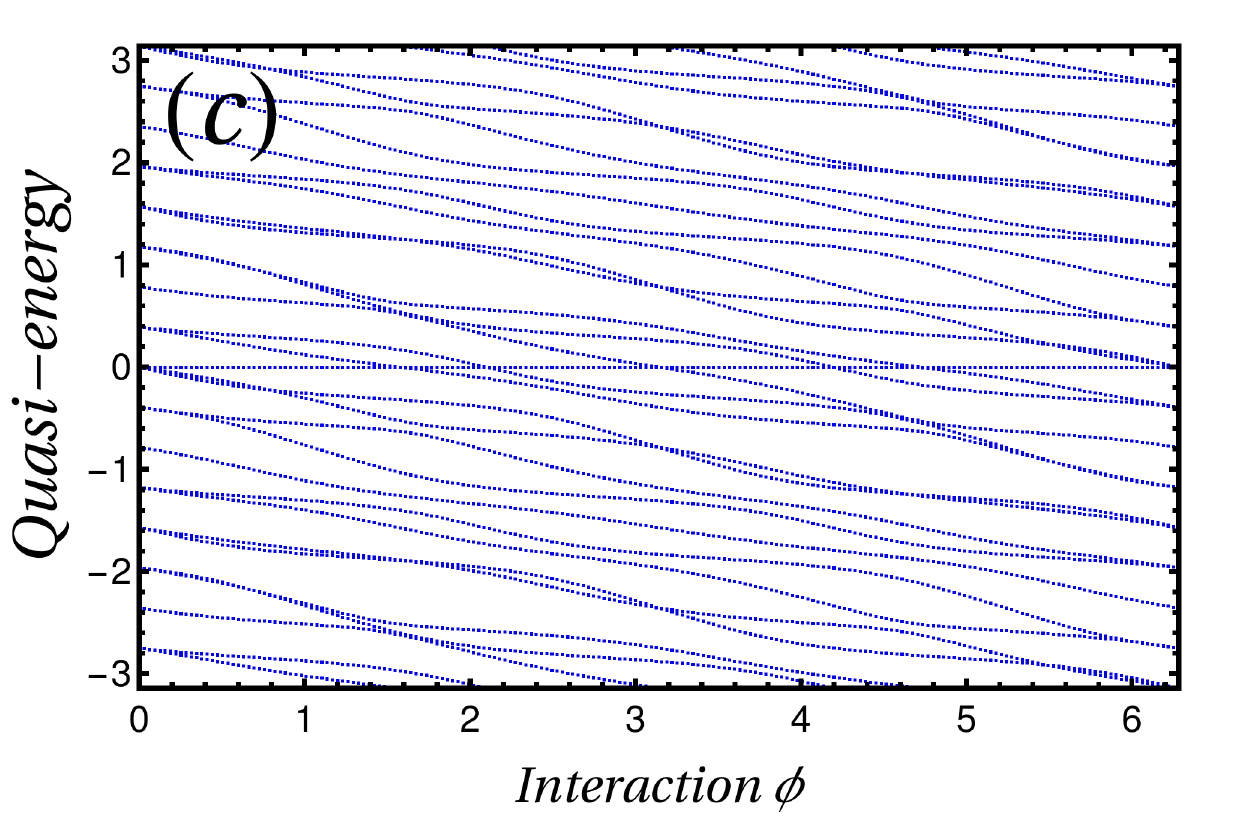}&  \includegraphics[width=0.25\textwidth]{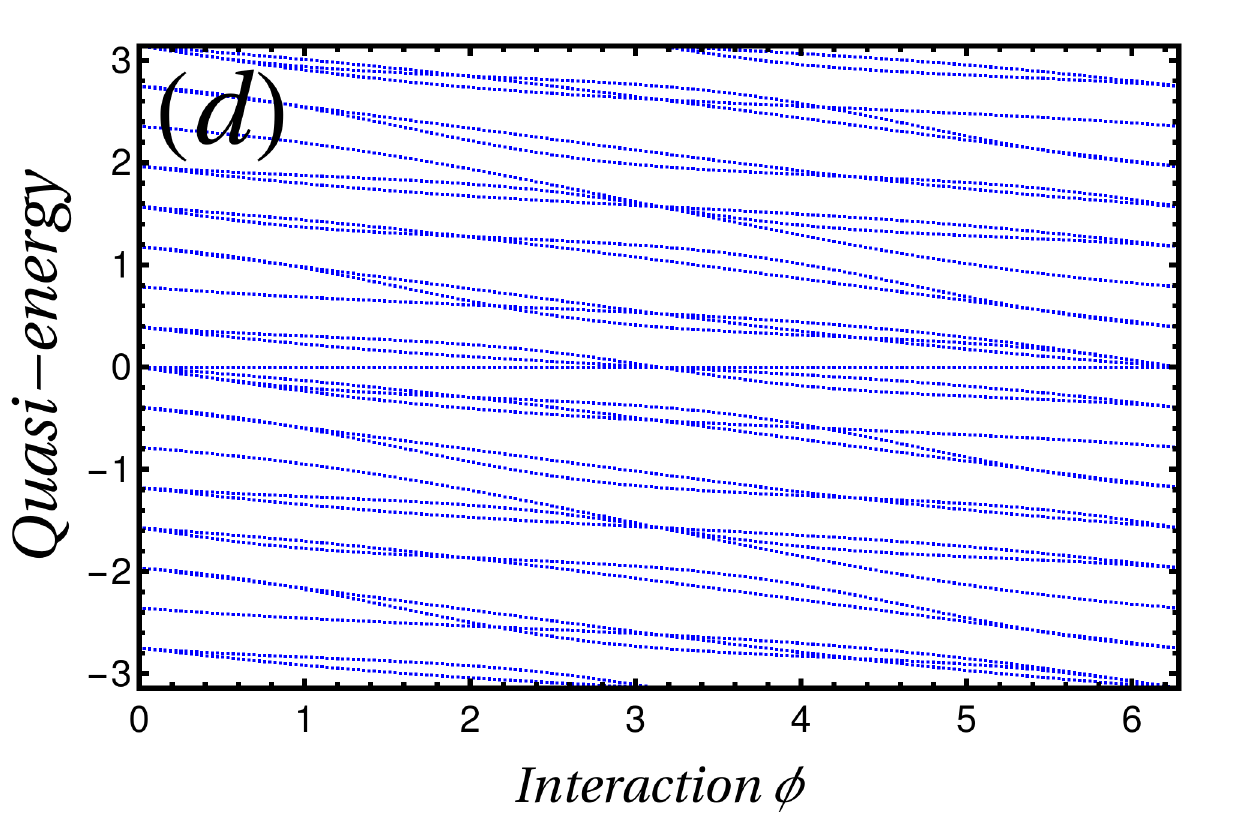}\\ \includegraphics[width=0.25\textwidth]{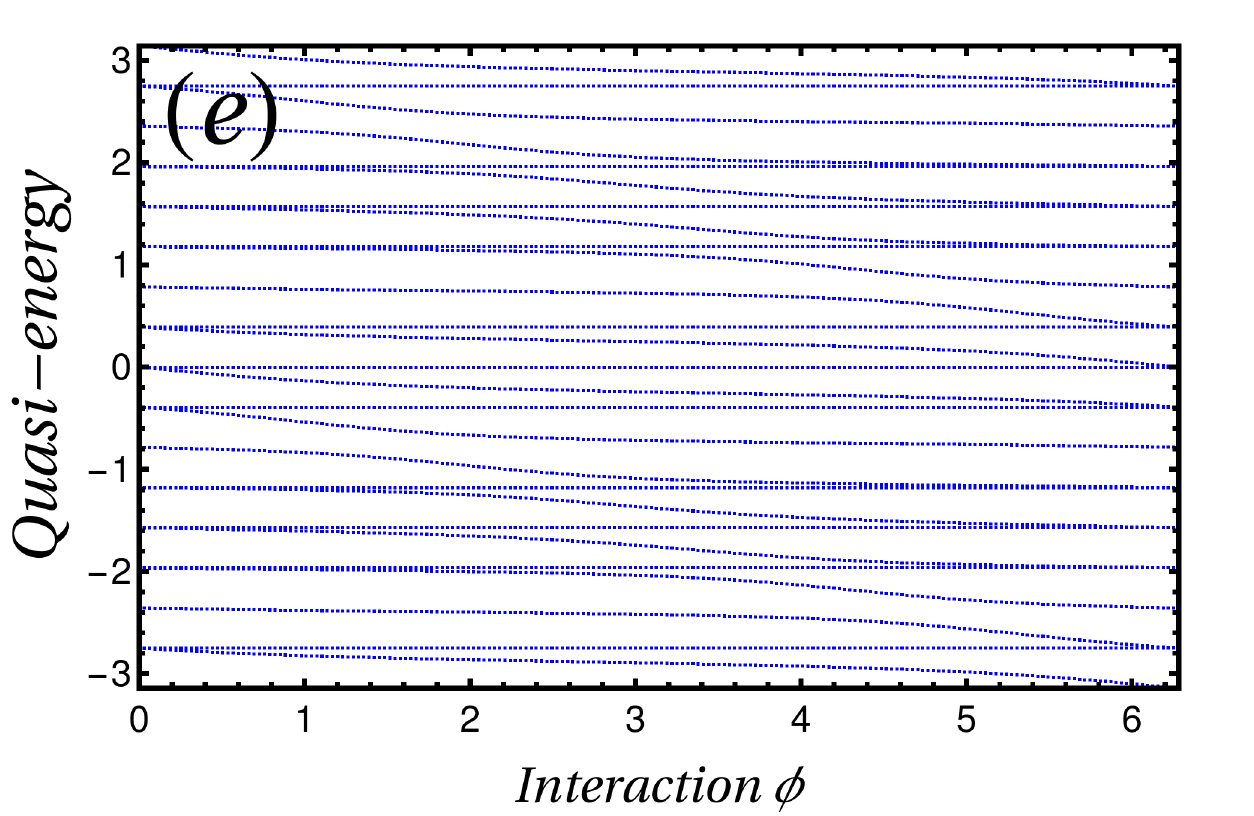}   &  \includegraphics[width=0.25\textwidth]{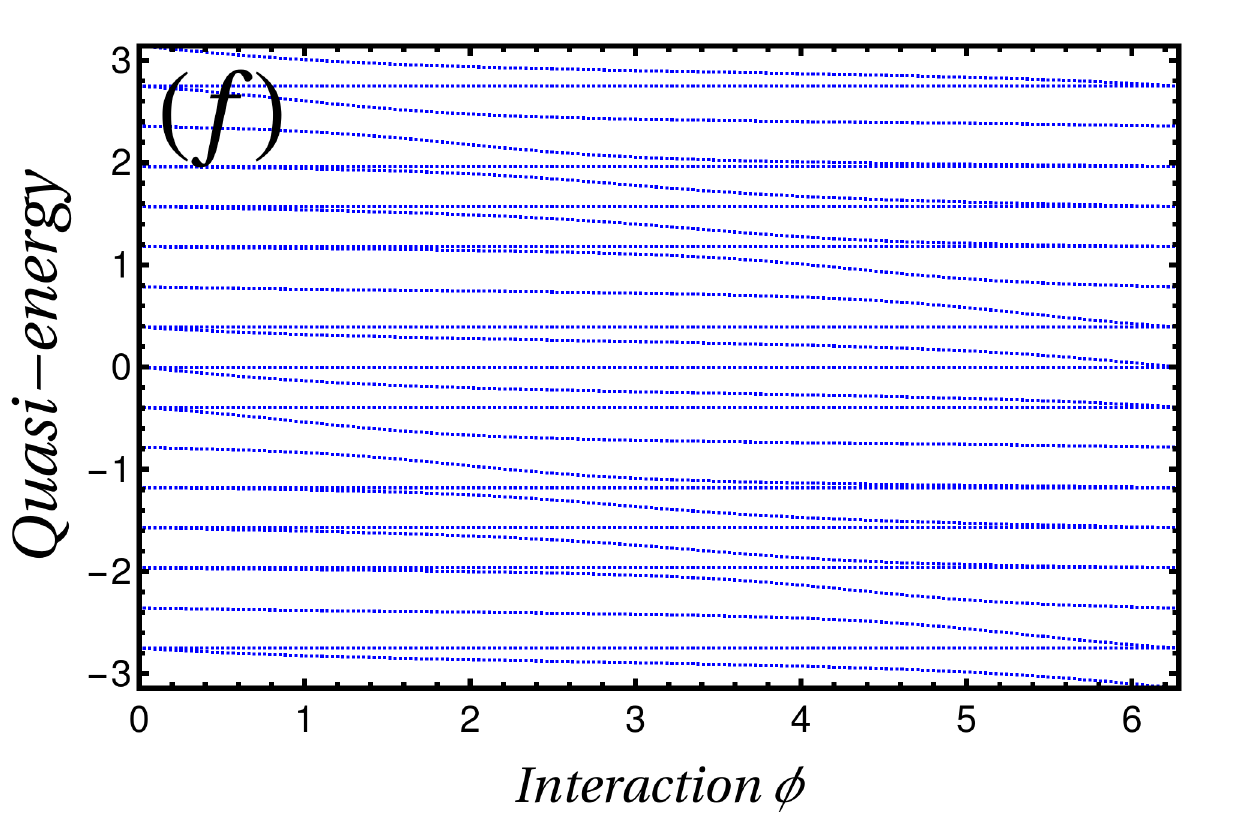}\\
\end{tabular}
   
  \caption{Quasi-energy spectra as a function of the interaction $\phi$ at the critical flux $f_c=1/2$ using the Grover coin on hub sites and the $U_2(\pi/4,0,0,0)$ coin on rim sites. The left panels: (a), (c), (e) (resp. right panels: (b), (d), (f) ) are the symmetric (resp. antisymmetric) parts of the spectrum. The top panels: (a) and (b) are the spectra associated to the subspace $\ket{\psi_0}$, the middle panels: (c) and (d) are associated with the subspace $\ket{\psi_1}$, and the bottom panels: (e) and (f) are associated with subspace $\ket{\psi_2}$.  The spectra associated to $\ket{\psi_0}_{S/AS}$ depend on $k_+$ (dispersive), they allow a motion of the center of mass and thus the destruction of the cages. The $4$ other spectra remain flat, they do not destroy the cages but their quasi-energies depend on the interaction parameter $\phi$.}
    \label{fig:spectreu}
\end{figure}

 As in the Hamiltonian version~\cite{Vidal2000}, we observe two families of energies depending on the interaction: dispersive and non-dispersive (or flat) bands. The dispersive bands are only due to  $\{\ket{\psi_0(k_+,\varepsilon,\varepsilon')}_{S/AS}\}_{\varepsilon,\varepsilon'}$ and break cages. On the contrary, the subspaces associated to $\{\ket{\psi_1(k_+,\varepsilon,\varepsilon')}_{S/AS}\}_{\varepsilon,\varepsilon'}$ and $\{\ket{\psi_2(k_+,\varepsilon,\varepsilon')}_{S/AS}\}_{\varepsilon,\varepsilon'}$ depend on the interaction $\phi$ but remain non-dispersive (see Figure~\ref{fig:spectreu}).\\

\par Therefore, the effect of interaction can lead to very different behaviors depending on the subspace considered. We conclude from the detailed analysis given in Appendix~\ref{ap:stability} that, on the one hand, subspaces generated by $\{\ket{n,\varepsilon,n+2,\varepsilon'}\}_{\varepsilon,\varepsilon'}$ for a given $n$ are stable under the interacting QW operator, i.e. $\forall \ket{\psi}\in \{\ket{n,\varepsilon,n+2,\varepsilon'}\}_{\varepsilon,\varepsilon'},\ W_{int}\ket{\psi}\in \{\ket{n,\varepsilon,n+2,\varepsilon'}\}_{\varepsilon,\varepsilon'} $. Subspaces generated by $\{\ket{n,\varepsilon,n+1,\varepsilon'}\}_{\varepsilon,\varepsilon'}$ or $\{\ket{n,\varepsilon,n-1,\varepsilon'}\}_{\varepsilon,\varepsilon'}$ are not stable under the action of the interacting QW operator. However, we can find a larger stable subspace $\{\ket{n,\varepsilon,n+1,\varepsilon'}\}_{\varepsilon,\varepsilon'}\bigoplus\{\ket{n+1,\varepsilon,n,\varepsilon'}\}_{\varepsilon,\varepsilon'}$. It mixes states of relative position $x_-=\pm1$ but still gives a trapped dynamics with a larger cage.

\par On the other hand, the minimum stable subspace for eigenvectors of relative position $x_-=0$ mixes all the different center of mass positions: $\underset{n}{\bigoplus}\{\ket{n,\varepsilon,n,\varepsilon'}\}_{\varepsilon,\varepsilon'}$, so that this subspace leads to the destruction of cages.

\par
\begin{figure}
    \centering
    \includegraphics[width=\linewidth]{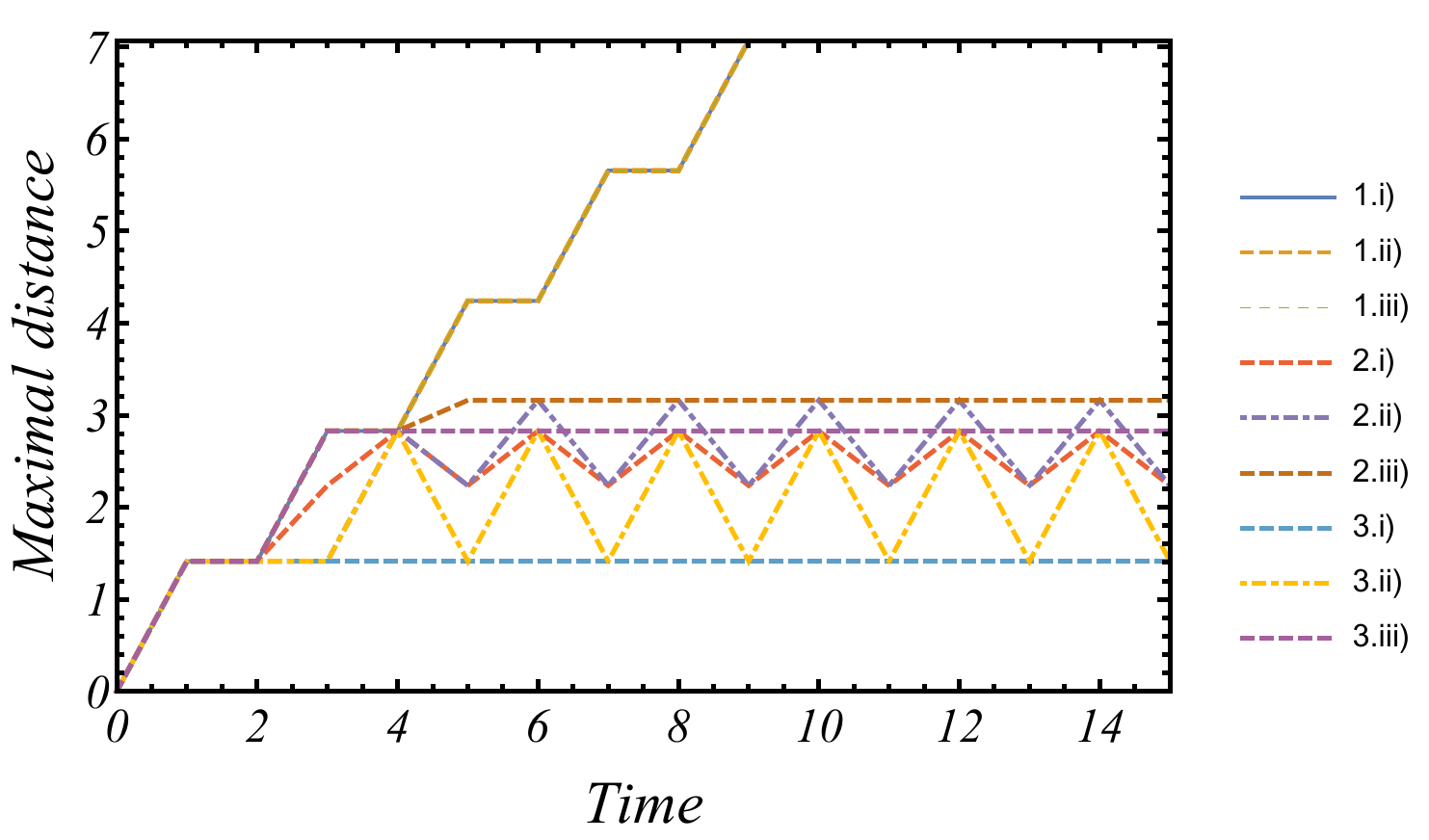}
    \caption{Maximal distance (in units of a cell length) that the walkers can reach in the 2-body Hilbert space versus time for 9 different configurations.}
    \label{fig:maxdist}
\end{figure}

 We numerically recover the analytical results obtained in Appendix~\ref{ap:stability} by computing the maximal distance in the Hilbert space (see Fig.~\ref{fig:maxdist}) that can be reached by the bound state from its initial site. We investigate this quantity for different initial states: 
\begin{enumerate}
    \item On the same site $a$ (the diagonal)
    \item On nearest-neighbour sites $a$ (the sub-diagonal)
    \item On next to nearest-neighbour sites $a$ (the second sub-diagonal)
\end{enumerate}
with an initial spin configuration:
\begin{enumerate}[label=\roman*)]
    \item $(\ket{1}+\ket{2}+\ket{3}+\ket{4})\otimes(\ket{1}+\ket{2}+\ket{3}+\ket{4})$
    \item $\ket{11}+\ket{44}$ 
     \item $\ket{11}+\ket{22}+\ket{33}+\ket{44}$ 
\end{enumerate}

\par The evolution of these states is obtained using a Grover coin at the critical flux $f_c=1/2$.  Based on the analysis made in Appendix~\ref{ap:stability}, we find that the initial state 1 breaks cages for any spin configuration i), ii), iii) while the two other initial states preserve cages. 

\par Subspaces to which initial states belong are summed up in the Table \ref{tab:ini}. The initial condition 1 for any initial spin configuration i), ii), iii) lives on  the subspace generated by $\{\ket{n,\varepsilon,n,\varepsilon'}\}_{\varepsilon,\varepsilon',n}$.  As a consequence, theses states break the cages.
On the contrary, the initial conditions 2 and 3 do not break the cage but their dynamics is affected by the interaction because they have only weight on, respectively, subspace $\{\ket{n,\varepsilon,n+1,\varepsilon'}\}_{\varepsilon,\varepsilon'}\bigoplus \{\ket{n+1,\varepsilon,n,\varepsilon'}\}_{\varepsilon,\varepsilon'}$  and $\{\ket{n,\varepsilon,n+2,\varepsilon'}\}_{\varepsilon,\varepsilon'}$. For initial conditions 2 and 3, the stable subspace on which the two QW live depends on the initial spin configuration. For initial condition $2$, the stable subspaces are:
\begin{enumerate}[label=\roman*)]
    \item $\{\ket{n,\varepsilon,n-1,\varepsilon'}\}_{\varepsilon,\varepsilon'}\bigoplus \{\ket{n-1,\varepsilon,n,\varepsilon'}\}_{\varepsilon,\varepsilon'}$
    \item $\bigoplus_{i=0}^1\left(\{\ket{n-i,\varepsilon,n+1-i,\varepsilon'}\}_{\varepsilon,\varepsilon'}\right.$ $\textrm{\hspace{.68cm}}\left.\oplus \{\ket{n+1-i,\varepsilon,n-i,\varepsilon'}\}_{\varepsilon,\varepsilon'}\right)$
    \item $\bigoplus_{i=-1}^1\left(\{\ket{n+i,\varepsilon,n+i-1,\varepsilon'}\}_{\varepsilon,\varepsilon'}\right.$ $\textrm{\hspace{.9cm}}\left.\oplus \{\ket{n+i-1,\varepsilon,n+i,\varepsilon'}\}_{\varepsilon,\varepsilon'}\right)$
\end{enumerate}
For the initial condition $3$, the stable subspace is the initial subspace indicated in the Table~\ref{tab:ini}.\\

\begin{table*}[ht]

\begin{tabular}{|c||c|c|c|}
\hline
initial state& a & b &c\\
\hline 
\hline
1: $(n,n)$ &$\{\ket{n,\varepsilon,n,\varepsilon'}\}_{\varepsilon,\varepsilon'}$&$\{\ket{n,\varepsilon,n,\varepsilon'}\}_{\varepsilon,\varepsilon'}\bigoplus\{\ket{n+1,\varepsilon,n+1,\varepsilon'}\}_{\varepsilon,\varepsilon'}$&$\bigoplus_{i=-1}^1\{\ket{n+i,\varepsilon,n+i,\varepsilon'}\}_{\varepsilon,\varepsilon'}$ \\
\hline
2: $(n,n-1)$ &$\{\ket{n,\varepsilon,n-1,\varepsilon'}\}_{\varepsilon,\varepsilon'}$&$\{\ket{n,\varepsilon,n-1,\varepsilon'}\}_{\varepsilon,\varepsilon'}\bigoplus\{\ket{n+1,\varepsilon,n,\varepsilon'}\}_{\varepsilon,\varepsilon'}$&$\bigoplus_{i=-1}^1\{\ket{n+i,\varepsilon,n+i-1,\varepsilon'}\}_{\varepsilon,\varepsilon'}$\\
\hline
3: $(n,n-2)$ &$\{\ket{n,\varepsilon,n-2,\varepsilon'}\}_{\varepsilon,\varepsilon'}$&$\{\ket{n,\varepsilon,n-2,\varepsilon'}\}_{\varepsilon,\varepsilon'}\bigoplus\{\ket{n+1,\varepsilon,n-1,\varepsilon'}\}_{\varepsilon,\varepsilon'}$&$\bigoplus_{i=-1}^1\{\ket{n+i,\varepsilon,n+i-2,\varepsilon'}\}_{\varepsilon,\varepsilon'}$\\
\hline
\end{tabular}

\caption{Subspaces on which the 9 different initial states live.} 
  \label{tab:ini}

\end{table*}

\begin{table}
 
\begin{tabular}{|c||c|c|c|}
\hline
initial state& a & b &c\\
\hline 
\hline
1: $(n,n)$ &$32L$&$32L$&$32L$ \\
\hline
2: $(n,n-1)$ &$64$&$128$&$192$\\
\hline
3: $(n,n-2)$ &$32$&$64$&$96$\\
\hline
\end{tabular}

\caption{ Dimension of the stable subspace of the 9 different initial states for a dimaond chain of length $L$.} 
  \label{tab:stable}
\end{table}

\par In Fig.~\ref{fig:maxdist}, we observe that configurations 1i), ii), iii) reach a maximal distance greater than the others. Their value is bounded $(L/\sqrt{2})$ due to the finite size of the chain $(L=10)$ but in principle this distance should keep increasing for an infinite chain because cages are broken. For configurations 2i),ii),iii) and 3i),ii),iii) maximal distances is bounded due to the presence of the cage. The value reached is correlated to the dimension of the stable subspace on which live the initial state (see Table~\ref{tab:stable}). The higher the dimension of the stable subspace, the greater the maximal distance.

\section{Conclusion \label{sec:conclusion}}

We studied different perturbations that may destroy the AB cages in a quantum walk on a diamond chain. Except one case (applying repeated measurements), these perturbations are chosen such that the unitary process (central for quantum walks) is maintained. This leads to focus on different types of phase disorder on the quantum coins. They are applied either on the hub sites and on the two rim sites; in the latter case, differences are observed for equal perturbations for the two rim sites (which do not introduce phase decoherence in the AB circuits), and for disymmetric perturbation. Using quench or dynamical disorder, or both, we observe a quite large set of behaviours in term of QW wavefunctions spreading: either a stopped diffusion ($\gamma=0$), a standard diffusion ($\gamma=1/2$) or a ballistic motion ($\gamma=1$). We also found a more surprising case of subdiffusive motion with non-trivial exponent $0<\gamma<1/2$ controlled by the interplay between static and dynamical disorders. We also study how an on-site interaction between two particle experiencing the same QW can also lead differences in spreading modes. In particular, we could observe a particular ballistic regime for a molecular bound state.

As perspectives and following Ref.~\cite{Ribeiro2004}, it would be worth studying the effect of static and dynamical aperiodic sequences on the AB cages in the QW on the diamond chain. There is a richness of such sequences ranging between quasi-periodic to disordered sequences (e.g. Fibonacci, double-period, Thue-Morse, etc). We expect to obtain anomalous diffusion in these cases as well, but probably with exponents $1/2<\gamma<1$. Another direction would be to test the robustness of QW AB cages also in 2D such as on the $\mathcal{T}_3$ lattice. We could also investigate the effect of a random magnetic field centered on the critical flux.

The transition between quantum to classical behavior comes from the interaction between the system and its environment. The way we recovered classical system using dynamical disorder or repeated measurement is an effective scheme allowing us to capture the main features of decoherence. In the hamiltonian framework, a more rigorous and general method to take into account environment effects relies on the density matrix formalism and the Lindblad equation~\cite{Lindblad1976}. It includes the description of mixed quantum states which make the connection between pure quantum and classical states. A similar approach is possible for QW systems and is known as Open Quantum Walks~\cite{Attal2012,Attal2012a}. Embedding the DC QW into a bath, it would certainly be interesting to compare the obtained dynamics with those described here in Secs.~\ref{sec:dyndis},~\ref{sec:measure} and ~\ref{sec:subdiffusion}.


\acknowledgements
We thank P. Ribeiro for sharing unpublished notes on the density matrix and O. B\'enichou and A. Barbier-Chebbah for useful discussions on anomalous diffusion.

\appendix

\section{Variance for dynamical disorder on rim sites}
\label{ap:diffusion}
\par In this appendix, we explain the numerical results of the dynamical disorder on the rim sites for a maximal disorder $\Delta \theta=2\pi$ (see Figure~\ref{fig:dynrim}-b and Sec.~\ref{sec:dynrimdis}) by using the formalism of the density matrix~\cite{Ribeiroprivate}. We will make two assumptions that we will verify numerically, the rest of the calculation is analytically exact.

 \par The evolution of the density matrix is given by $\rho(t+1)=U(t).\rho(t).U^\dagger(t)$. Let us first compute the quantum walk operator. We suppose that the quantum walker is initially localized on a hub site in a certain internal configuration. In order to have only $4$ states per cell (instead of $8$), we are interested in the evolution of the quantum walk operator at even times when the walker is on the hub sites. We compute the aggregate quantum walk operator on two time steps $\hat{W}(t)=\hat{U}(2t+1).\hat{U}(2t)$. The calculation is done in the case of a Grover coin $G_4$ with magnetic flux $f=1/2$, the system without disorder is thus caging. We find
\begin{eqnarray}
\hat{W}(t)=\sum_{i=-\infty}^{+\infty}&&
\begin{pmatrix}
0&c_+&-c_-&0\\
c_+&0&0&-c_-\\
0&c_-&-c_+&0\\
c_-&0&0&-c_+
\end{pmatrix}\otimes\ket{i,a}\bra{i,a}\nonumber\\
+&&
\begin{pmatrix}
0&0&0&0\\
0&-s_-&s_+&0\\
0&0&0&0\\
0&-s_+&s_-&0
\end{pmatrix}\otimes\ket{i+1,a}\bra{i,a}\nonumber\\
+&&
\begin{pmatrix}
s_-&0&0&-s_+\\
0&0&0&0\\
s_+&0&0&-s_-\\
0&0&0&0
\end{pmatrix}\otimes\ket{i-1,a}\bra{i,a}\nonumber
\end{eqnarray}
where $\ket{i,a}$ denotes the hub site $a$ on the cell index $i$, the matrices represent the internal spin space  written in the basis $\{\ket{R^+},\ket{L^+},\ket{R^-},\ket{L^-}\}$and $c_+$, $c_-$, $s_+$ and $s_-$ are shortcut notations for $c_+=\frac{\cos\theta_b+\cos\theta_c}{2}$, $c_-=\frac{\cos\theta_b-\cos\theta_c}{2}$, $s_+=\frac{\sin\theta_b+\sin\theta_c}{2}$ and $s_-=\frac{\sin\theta_b-\sin\theta_c}{2}$ where $\theta_b$ and $\theta_c$ are chosen independently at each time step in the box $[\theta_0-\Delta\theta/2,\theta_0+\Delta\theta/2 ]$. From the solution of the QW we deduce the evolution of the density matrix which is $\rho_W(t+1)=W(t).\rho_W(t).W^\dagger(t)$. The problem is (space) translationally invariant, one can use the generating function (similar to the discrete Fourier transform): 
\begin{equation}
    P^{\alpha\beta}(t,w,z)=\sum_{i,j} \rho^{\alpha\beta}_{i,j}(t)w^{\frac{i+j}{2}}z^{i-j}
\end{equation} where $\alpha,\beta=R^+,R^-,L^+,L^-$ and $\rho^{\alpha\beta}_{i,j}(t)=\bra{i,a}\rho^{\alpha\beta}_W(t)\ket{j,a}$. Thanks to the density matrix hermiticity $(\rho^{\alpha\beta}_{i,j})^\dagger(t)=(\rho^{\beta\alpha}_{j,i})^*(t)=\rho^{\alpha\beta}_{i,j}(t)$, the corresponding  generating function satisfies the relation $P^{\alpha\beta}(t,w,z)=(P^{\beta\alpha}(t,w,1/z))^*$. In particular for diagonal terms, there is no dependence in $z$. We therefore have 16 equations (but only 10 independent ones because of the last relation) of the form 
\begin{equation}
P^{\alpha\beta}(t+1,w,z)=\sum_{\alpha',\beta'}f_{\alpha,\beta,\alpha',\beta'}(t,w,z)P^{\alpha'\beta'}(t,w,z)
\label{eq:genfunction}
\end{equation} where $f_{\alpha,\beta,\alpha',\beta'}(t,w,z)=\frac{a}{w}+\frac{b}{z^2}+\frac{c}{\sqrt{w} z}+d \frac{z}{\sqrt{w}}+e+f\frac{\sqrt{w}}{z}+g \frac{z}{\sqrt{w}}+hz^2+i w$ where $a,b,c,d,e,f,g,h,i$ depend on the indices $\alpha',\beta',\alpha,\beta$ and the random variables $\theta_b(2t+1)$ and $\theta_c(2t+1)$. 
Using generating functions, the quantum walk's variance can be written:
\begin{align}
    \sigma^2(t)&\equiv\sum_i \sum_{\alpha}i^2|\bra{i,a,\alpha}\ket{\psi(t)}|^2\nonumber\\
    &-\left(\sum_i \sum_{\alpha}i|\bra{i,a,\alpha}\ket{\psi(t)}|^2\right)^2 \label{eq:sigmagen1}\\
&=\partial_w^2 \text{Tr}(P^{\alpha\beta}(t,1,0))+\partial_w \text{Tr}(P^{\alpha\beta}(t,1,0))\nonumber\\
&-\left(\partial_w \text{Tr}(P^{\alpha\beta}(t,1,0))\right)^2
    \label{eq:sigmagen2}
\end{align}
where $\alpha=R^+,L^+,R^-,L^-$ and the trace is on the internal spin space.
\par From here, we make two hypotheses that we checked numerically but for which the formal proof is not given. In Sec.~\ref{sec:dyndis}, we are interested in the standard deviation averaged over the disorder i.e. we first take the square root of the variance and then we average it. We will see later that $P^{\alpha\beta}(t,1,0)$ is easily computable once the average over the disorder is taken. However, for the calculation of the standard deviation the square root function is not linear, so it is not possible a priori to invert both operations. One will thus be interested thereafter in calculating the variance averaged over the disorder and not the standard deviation. Moreover, we checked numerically that $\sqrt{\overline{\sigma^2}}\simeq\overline{\sigma}$ up to $10^{-2}$.
We now average the variance over the disorder:
\begin{align}
    \overline{\sigma^2(t)}&=\overline{\partial_w^2 \text{Tr}(P^{\alpha\beta}(t,1,0))}+\overline{\partial_w \text{Tr}(P^{\alpha\beta}(t,1,0))}\nonumber\\
    &-\overline{\left(\partial_w \text{Tr}(P^{\alpha\beta}(t,1,0))\right)^2}\label{eq:sigm1}\\
    &=\partial_w^2 \text{Tr}(\overline{P^{\alpha\beta}}(t,1,0))+\partial_w \text{Tr}(\overline{P^{\alpha\beta}}(t,1,0))\nonumber\\
    &-\overline{\left(\partial_w \text{Tr}(P^{\alpha\beta}(t,1,0))\right)^2}\label{eq:sigm2}
\end{align}
For the first two terms of the equation~\eqref{eq:sigm1} we can exchange the average over the disorder and the derivatives along $w$ as well as the trace because they are linear operators, which gives the equation~\eqref{eq:sigm2}. For the last term this is a priori not possible because of the square.
The first two terms correspond to the second order moment of the wave function distribution:
\begin{align}
    \mu_2(t)&\equiv\sum_i \sum_{\alpha}i^2|\bra{i,a,\alpha}\ket{\psi(t)}|^2\nonumber\\
    &=\partial_w^2 \text{Tr}(P^{\alpha\beta}(t,1,0))+\partial_w \text{Tr}(P^{\alpha\beta}(t,1,0))\\
 \overline{\mu_2(t)}   &=\partial_w^2 \text{Tr}(\overline{P^{\alpha\beta}}(t,1,0))+\partial_w \text{Tr}(\overline{P^{\alpha\beta}}(t,1,0))
\end{align}
The last term corresponds to the average position of the squared wave function averaged over the disorder which will be written:
\begin{equation}
 \overline{\mu_1^2(t)}= \overline{\left(\partial_w \text{Tr}(P^{\alpha\beta}(t,1,0))\right)^2}
\end{equation}
We obtain:
\begin{equation}
    \overline{\sigma^2(t)}= \overline{\mu_2(t)}-\overline{\mu_1^2(t)}
    \label{eq:sigm3}
\end{equation}
The second hypothesis consists in inverting the square of the mean position of the walker and the average over the disorder:
\begin{align}
    \overline{\sigma^2(t)}&\simeq \overline{\mu_2(t)}-\overline{\mu_1(t)}^2\nonumber\\
    &=\partial_w^2 \text{Tr}(\overline{P^{\alpha\beta}}(t,1,0))+\partial_w \text{Tr}(\overline{P^{\alpha\beta}}(t,1,0))\nonumber\\&-\left(\partial_w \text{Tr}(\overline{P^{\alpha\beta}}(t,1,0))\right)^2\label{eq:sigmapprox}
\end{align}

We checked numerically that the difference between equations~\eqref{eq:sigm3} and \eqref{eq:sigmapprox} is $10^{-11}$. 
\par We are now interested in computing $\overline{P^{\alpha\beta}}(t,w,z)$. Using equation~\eqref{eq:genfunction}, the function $f_{\alpha,\beta,\alpha',\beta'}(t,w,z)$ depends only on the random variables $\theta_b$ and $\theta_c$ at time $2t+1$, then $P^{\alpha\beta}(t,w,z)$ depends only on these variables but at time $2t'+1$ with $t'<t$. The two quantities are thus independent and the functions $\overline{f_{\alpha,\beta,\alpha',\beta'}}(w,z)$ once averaged over the disorder are independent of time. The time equation averaged over the disorder then becomes:
\begin{equation}
  \overline{P^{\alpha\beta}}(t+1,w,z)=\sum_{\alpha',\beta'}\overline{f_{\alpha,\beta,\alpha',\beta'}}(w,z)\overline{P^{\alpha'\beta'}}(t,w,z)  .
\end{equation}
\par The problem is now time translation invariant, so we can use the generating function on the time variable (similar to a discrete Laplace transform):
\begin{equation}
   Q^{\alpha\beta}(s,w,z)=\sum_{t=0}^\infty \overline{P^{\alpha\beta}}(t,w,z)s^t 
\end{equation} and the equation on $Q$ can be written:
\begin{eqnarray}
Q^{\alpha\beta}(s,w,z)&=&P^{\alpha\beta}(0,w,z)\nonumber\\
&+&s\sum_{\alpha',\beta'}\overline{f_{\alpha,\beta,\alpha',\beta'}}(w,z)Q^{\alpha'\beta'}(s,w,z)\nonumber
\end{eqnarray} 
We are then reduced to solve a linear problem of algebra. For the sake of simplicity, we take a maximal disorder $\Delta\theta=2\pi$. In this particular case, we remark that the unknowns $Q^{R^+R^+}(s,w,z)$, $Q^{L^+L^+}(s,w,z)$, $Q^{R^-R^-}(s,w,z)$ and $Q^{L^-L^-}(s,w,z)$ form a subset we note $Q^{R^+R^+}(s,w,z)=Q^{R^+}$, $Q^{L^+L^+}(s,w,z)=Q^{L^+}$, $Q^{R^-R^-}(s,w,z)=Q^{R^-}$ and $Q^{L^-L^-}(s,w,z)=Q^{L^-}$.
$$
\left\{
    \begin{array}{l}
    Q^{R^+}=|a_0|^2+\frac{s}{4w}\left[Q^{R^+}+Q^{L^-}+(Q^{L^+}+Q^{R^-})w\right]\\
    Q^{L^+}=|b_0|^2+\frac{s}{4}\left[Q^{R^+}+Q^{L^-}+(Q^{L^+}+Q^{R^-})w\right]\\
    Q^{R^-}=|c_0|^2+\frac{s}{4w}\left[Q^{R^+}+Q^{L^-}+(Q^{L^+}+Q^{R^-})w\right]\\
    Q^{L^-}=|d_0|^2+\frac{s}{4}\left[Q^{R^+}+Q^{L^-}+(Q^{L^+}+Q^{R^-})w\right]
    \end{array}
\right.
$$
where $(a_0,b_0,c_0,d_0)$ is the initial spin configuration in the basis $(\ket{R^+},\ket{L^+},\ket{R^-},\ket{L^-})$ located on the site $a$ of index cell $i=0$. The initial state is normalised $|a_0|^2+|b_0|^2+|c_0|^2+|d_0|^2=1 $. Taking the sum over the 4 solutions, we find $\text{Tr}(\overline{Q^{\alpha\beta}}(s,w,z))$. We deduce :
\begin{align}
\sum_t\overline{\mu_1(t)}s^t &=\partial_w\text{Tr}(Q^{\alpha\beta}(s,1,0))\nonumber\\
 &=\frac{s(|a_0|^2-|b_0|^2-|c_0|^2+|d_0|^2}{2(1-s)} \nonumber\\
 &=\frac{s(|a_0|^2-|b_0|^2-|c_0|^2+|d_0|^2} {2}\sum_{p=0}^\infty s^p \nonumber\\
&=\frac{|a_0|^2-|b_0|^2-|c_0|^2+|d_0|^2} {2}\sum_{p=1}^\infty s^p  \\
\sum_t\overline{\mu_2(t)}s^t&=\partial_w^2  \text{Tr}(Q^{\alpha\beta}(s,1,0))+\partial_w\text{Tr}(Q^{\alpha\beta}(s,1,0))\nonumber\\
&=\frac{s} {2(1-s)^2} =\f{s}{2}\sum_{m,n=0}^\infty s^{n+m}=\f{s}{2}\sum_{p}(p+1) s^{p}\nonumber\\
&=\sum_{p=1}^\infty \f{p}{2} s^p
\end{align}
where in the second last line $p=n+m$.
\par Here, the equation~\eqref{eq:genfunction} is computed for the quantum walk operator aggregated over two time steps $\hat{W}(t)$. We must therefore divide the time by two to get back to the original time unit and we find:
\begin{equation}
    \sum_t\overline{\mu_2(t)}s^t=\sum_{t=1}^\infty \f{t}{4} s^t
\end{equation}
We obtain for $t>1$:
\begin{equation}
\overline{ \sigma^2(t)}\simeq\f{t-  \left(|a_0|^2-|b_0|^2-|c_0|^2+|d_0|^2\right)^2}{4}
\end{equation} 

\par It is clear from this calculation that the variance is linear in time. The calculation was possible thanks to the translation invariance (of space) of the problem and the independence of the random variables at different times.  In the large time limit, we finally find the numerical fit result of Figure~\ref{fig:dynrim}-b:
\begin{equation}
    \overline{\sigma(t)}\simeq\sqrt{\overline{\sigma^2}(t)}\underset{t\to\infty}{\sim}\f{\sqrt{t}}{2}
\end{equation}
\section{Construction of the second lattice in the 2-body Hilbert space}
\label{ap:lattice2}
The construction of the second lattice is composed of the remaining sites $(a,b)$, $(a,c)$, $(b,a)$ and $(c,a)$. $(a,b)$ and $(a,c)$  with $(n,m)$ cell indices are connected to $(b,a)$ and $(c,a)$ sites, with cell indices  ($(n,m)$, $(n-1,m)$, $(n,m+1)$ and $(n-1,m+1)$. Similarly, $(b,a)$ and $(c,a)$  with $(n,m)$ cell indices are connected to $(a,b)$ and $(a,c)$ sites, with cell indices  ($(n,m)$, $(n+1,m)$, $(n,m-1)$ and $(n+1,m-1)$.
\begin{figure}[h]
   \includegraphics[width=0.5\textwidth]{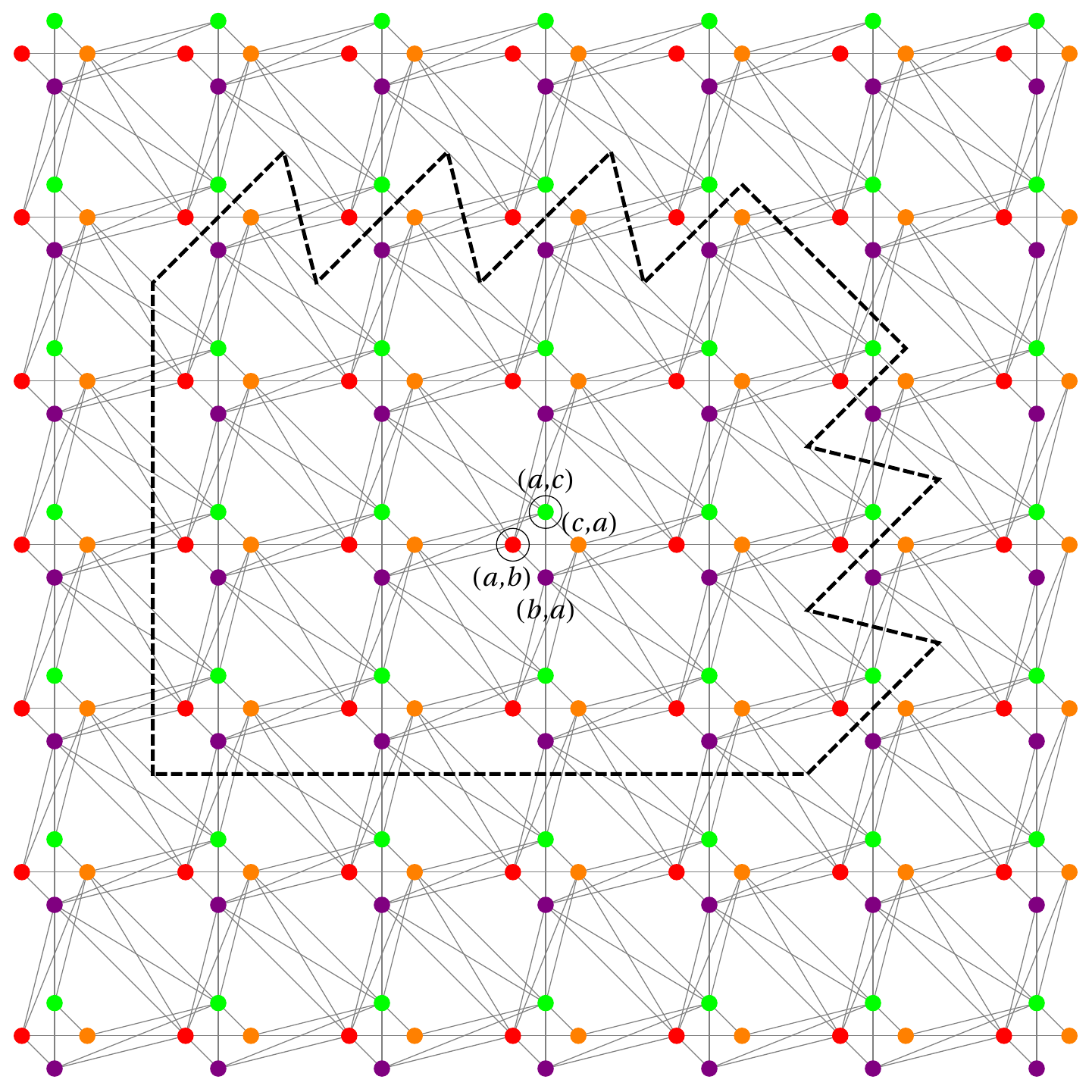}
  \caption{A piece of the second sublattice of the 2-body DC QW with sites of
type $(a,b)$, $(a,c)$, $(b,a)$, $(c,a)$ and $(c,c)$ (colored in red, green, purple and orange). The black dashed
rectangle indicates the maximal extension of a cage (at $f_c$),
for an initial state localized at the circled $(a,b)$ or $(a,c)$. The internal states are not represented.}
    \label{fig:ssres2}
\end{figure}

\section{Stability of subspaces under the interacting QW operator}
\label{ap:stability}
\par In this Appendix we give an explanation of the stability of subspaces $\{\ket{n,\varepsilon,n+2,\varepsilon'}\}_{\varepsilon,\varepsilon'}$, $\{\ket{n,\varepsilon,n+1,\varepsilon'}\}_{\varepsilon,\varepsilon'}\bigoplus\{\ket{n+1,\varepsilon,n,\varepsilon'}\}_{\varepsilon,\varepsilon'}$ and $\{\ket{n,\varepsilon,n,\varepsilon'}\}_{\varepsilon,\varepsilon',n}$ under the interaction. We refer back to the eigenvectors of the 1-body problem (see Figure \ref{fig:eigenvectors})


\par For states $\ket{n,\varepsilon_1,n+2,\varepsilon_2}$, quantum walkers only overlap on the $(a,a)$  sites of index cell $(n+1,n+1)$ corresponding to a  hub site at the boundary of the 1-body eigenstate for each quantum walker. This is the right hub site for the eigenvector of the first quantum walker $\ket{n,\varepsilon_1}$ (with amplitude probability proportional to $\delta$ on Figure \ref{fig:eigenvectors}) and the left hub site for the eigenvector of the second quantum walker $\ket{n+2,\varepsilon_2}$ (with amplitude probability proportional to $\alpha$ on Figure \ref{fig:eigenvectors}). States on this site are eigenvectors of the hub coin. Then the interaction multiplies this site by a global phase. As a consequence, these states on the $(a,a,n+1,n+1)$ site are still eigenvectors and their evolution won't leak out from its former cage. However, the whole state $\ket{n,\varepsilon_1,n+2,\varepsilon_2}$ is not an eigenstate anymore because of the overall interacting phase taken on the hub site $(n+1,n+1)$. After one time step it becomes a linear combination of the family $\{\ket{n,\varepsilon,n+2,\varepsilon'}\}_{\varepsilon,\varepsilon'}$. The same reasoning works for $\ket{n,\varepsilon_1,n-2,\varepsilon_2}$.

\par For states $\ket{n,\varepsilon_1,n,\varepsilon_2}$ associated to $\ket{\psi_0}$ quantum walkers overlap on 7 sites: 3 $(a,a)$ sites of index cells $(n-1,n-1)$, $(n,n)$ and $(n+1,n+1)$ and 2 $(b,b)$ and 2 $(c,c)$ sites of index cells $(n-1,n-1)$ and $(n,n)$. The interaction multiplies these sites by a global phase $e^{i\phi}$. However, the $(b,c)$ and $(c,b)$ sites are not affected by this phase. This asymmetry in the rim-rim sites (i.e. between $(b,b)$, $(c,c)$ in one hand and $(b,c)$ and $(c,b)$ sites in the other hand) leads to the destruction of cages. Indeed, for instance the rim-rim sites of the cell $(n,n)$, lead to the vector $\ket{L^-,L^-}$  on the $(a,a,n+1,n+1)$ sites after one time step QW without interaction. This vector is an eigenvector of the coin on $(a,a)$ sites. Therefore, it is bounced back inside the cage. When we turn on the interaction a part of the wavefunction on the $(a,a,n+1,n+1)$ sites coming from those rim-rim sites  will be proportional to the vectors $\ket{L^+,L^+}$. This state is not an eigenvector of the hub coin. At the next time step it will populate the state $\ket{n+1,\varepsilon'_1,n+1,\varepsilon'_2}$, leading to the leakage of quantum walkers outside of its former cage. 

\par For states $\ket{n,\varepsilon_1,n+1,\varepsilon_2}$, quantum walkers overlap onto 2 $(a,a)$ sites of index cells $(n,n)$ and $(n+1,n+1)$ and the $(b,b)$ and $(c,c)$ sites of index cell $(n,n)$. A priori, because of this asymmetry on the rim-rim sites, cages should not stand anymore. However, it appears that in this case, the leaking part is exclusively proportional to state on $(a,a)$ sites of wavefunctions $\{\ket{n,\varepsilon,n-1,\varepsilon}\}_{\varepsilon,\varepsilon'}$. The other way around, the interacting part of $\ket{n,\varepsilon_1,n-1,\varepsilon_2}$ recovers the family $\{\ket{n,\varepsilon,n+1,\varepsilon}\}_{\varepsilon,\varepsilon'}$. As a consequence, subspaces $\{\ket{n,\varepsilon,n+1,\varepsilon'}\}_{\varepsilon,\varepsilon'}\bigoplus\{\ket{n+1,\varepsilon,n,\varepsilon'}\}_{\varepsilon,\varepsilon'}$ remain stable for any given $n$, therefore the cage remains in this situation but is larger.\\

\bibliographystyle{apsrev4-2} 
\bibliography{ABQWbroken}
\end{document}